\documentclass[journal]{IEEEtran}
\usepackage{cite}

\usepackage{color}
\usepackage{enumerate}
\usepackage{xspace}

\makeatletter
\DeclareRobustCommand\onedot{\futurelet\@let@token\@onedot}
\def\@onedot{\ifx\@let@token.\else.\null\fi\xspace}

\makeatother

\newcommand{\includephoto}[1]{\includegraphics[width=1in,height=1.25in,clip,keepaspectratio]{#1}}
\newif\ifdebug
\ifdebug

\newcommand{\chenupdated}[1]{\hfill \textcolor{blue}{Chen Updated@#1}}

\newcommand{\liupdated}[1]{\hfill \textcolor{red}{Li Updated@#1}}

\else

\newcommand{\chenupdated}[1]{}

\newcommand{\liupdated}[1]{}

\newcommand{\qiaoupdated}[1]{}
\fi
\newif\ifFULLPAPER
\ifFULLPAPER
\newcommand{\full}[1]{#1}
\else
\newcommand{\full}[1]{}
\fi
\newif\ifTII
\ifTII
\newcommand{\TII}[1]{grey}{MENET: #1}
\else
\newcommand{\TII}[1]{}
\fi

\ifCLASSINFOpdf
  \usepackage[pdftex]{graphicx}
  \graphicspath{{img/}}
  \DeclareGraphicsExtensions{.pdf,.jpeg,.png}
\else
  \usepackage[dvips]{graphicx}
  \graphicspath{{img/}}
  \DeclareGraphicsExtensions{.eps}
\fi
\usepackage{amsmath, amsthm}
\usepackage{amsfonts}
\usepackage{amssymb}
\usepackage{makecell}
\usepackage{multirow}
\usepackage{enumitem}
\usepackage{bm}
\usepackage{indentfirst}

\usepackage{algorithmic}
\usepackage{algorithm}
\usepackage{pifont}

\usepackage{stfloats}
\usepackage{url}
\usepackage{hyperref}

\usepackage{subfig}
\usepackage{color}
\usepackage{bbding}
\usepackage[dvipsnames]{xcolor}
\usepackage{pifont}

\makeatletter

\newcommand{\Rmnum}[1]{\expandafter\@slowromancap\romannumeral #1@}
\makeatother

\begin{document}
%

\title{Efficient cross-domain multi-factor authentication}

%

\author{Yongjiao Li, Zhenlei Wang*, Zhu Cao*
\thanks{Y. Li and Z. Wang are with Key Laboratory of Smart Manufacturing in Energy Chemical Process, Ministry of Education, East China University of Science and Technology, China. (e-mail: wangzhen\_l@ecust.edu.cn)}
\thanks{Z. Cao is with College of Electronics and Information Engineering, Tongji University, Shanghai 201804, China, and Shanghai Research Institute for Intelligent Autonomous Systems, Tongji University, Shanghai 201210, China. (e-mail: caozhu@tongji.edu.cn)}

\thanks{\emph{*Corresponding authors: Zhenlei Wang and Zhu Cao.}}}

\maketitle

\begin{abstract}
Secure cross-domain authentication is crucial for protecting information exchange across terminals in different security domains. However, existing cross-domain authentication models suffer from inflexibility, high storage and computational consumption. In response to these problems, a cross-domain multi-factor authentication protocol with improved efficiency (CDAuth) is proposed. CDAuth employs a universally applicable multi-factor authentication approach, integrating pseudo-identity, attribute concealment, and biometric hiding. The authentication information sets can be freely added or modified to accommodate various access scenarios. This boosts privacy, flexibility, and applicability compared to traditional methods. Furthermore, multi-edge servers as Certification Authorities (CAs) store only the union of terminal attributes and a verification token per user in their subdomains, instead of each user’s attribute and biometric sets as in traditional schemes, significantly reducing storage consumption. Additionally, replacing the bilinear pairing operations with the native group operations, while maintaining security, significantly reduces computational consumption. Our analysis shows that CDAuth can resist common security attacks and is more efficient than the previous schemes.

\end{abstract}


%
\IEEEpeerreviewmaketitle

\section{Introduction}
\IEEEPARstart{T}{he} rapid growth of digitization and networking is connecting more devices and driving the expansion of smart cities, homes, and manufacturing \cite{saleem2024cost,9770364}. This increased connectivity enhances information exchange and societal progress but also raises significant threats to security and privacy \cite{saleem2025provably,jiang2024cross}. Effective authentication of the numerous devices is crucial for protecting privacy and ensuring network security. Emerging technologies like AI, big data, and 5G facilitate cross-regional and cross-departmental data sharing  \cite{10556609,10599865}. For example, in medical alliance systems, secure connections allow authorized users to access medical resources from other institutions. Terminals must first authenticate within their hospital before accessing external information, facilitating cross-domain cooperation and improving the efficiency of medical services \cite{xue2022cross}. However, if different encryption settings or PKI systems are used without proper identification and verification, access may fail. Therefore, cross-domain authentication is a critical issue.

Most cross-domain authentication systems \cite{article,10816446,shahidinejad2024all} rely on Public Key Infrastructure (PKI) \cite{cooper2021internet}, where certificate authorities (CAs) manage digital certificates. However, certificate management is costly, and revocation is challenging. Identity-Based Cryptography (IBC) \cite{11049050,9200782} simplifies this by using identity (ID) for authentication, removing the need for certificates. Yet, single-ID authentication suffers from weaker security and privacy risks \cite{alladi2020harci}, and IBC schemes often lack scalability. Attribute-based authentication \cite{11048743,huang2022blockchain} provides better flexibility and scalability but remains vulnerable to theft and tampering. Biometric authentication, being portable and unique, mitigates these issues, but current schemes \cite{liu2022low,meng2024bioinspired} mainly emphasize accuracy and efficiency while neglecting the privacy of biometric data. Multi-factor authentication \cite{ali2024enhanced,shukla2022novel} improves security by combining ID and biometric, typically on smart cards, but introduces the burden of managing smart cards, the high system overhead and the authentication granularity remains coarse and inflexible. Although \cite{tian2024identity,9785665} improve anonymity and privacy, challenges persist, including device vulnerabilities, dependence on universal factors, limited flexibility, and high computational and storage costs.

The large-scale connection of terminals requires extensive terminal user authentication. Relying on a single authority for distributing attributes, biometrics, and ID-based authentication, as well as managing keys, places a heavy burden on that authority. If compromised, the entire system collapses. To address this, multi-authority authentication is introduced\cite{10384713,10089843}, multiple authorities collaborate for intra-domain and cross-domain authentication. Some cross-domain authentication schemes introduce blockchain technology to enable multi-authority authentication \cite{10919182,10400828}, using single or multiple chains along with smart contracts and consensus mechanisms for decentralized authentication. In the approaches  \cite{10384713,fang2024decentralized}, edge/fog computing is integrated with blockchain to improve response times and reduce single-point failures. However, blockchain generation demands substantial time and effort, while blockchain read/write operations during authentication consume significant performance resources. 

Alternatively, some schemes deploy a central server in each domain for local authentication, with inter-server collaboration enabling cross-domain access \cite{10089843,penuelas2024revocable}. This distributed approach improves authentication efficiency and speed without the high energy consumption of traditional blockchain methods \cite{xue2022cross,10384713}. However, it lacks the immutability and traceability advantages of blockchain, and the legitimacy and security of the central servers are not fully addressed. Additionally, while single ID authentication in these schemes improves efficiency, it remains vulnerable to network attacks.

In this paper, we propose an efficient multi-authority cross-domain multi-factor authentication scheme called CDAuth. It uses an edge-cloud framework, where the cloud acts as the CA managing and authenticating edge servers, while edge servers near users handle local authentication and management. CDAuth improves on previous research by addressing its limitations and strengthening cross-domain authentication and information security. The main contributions are as follows.

(1) An enhanced multi-factor authentication method is proposed, combining ID, attributes, and biometrics to enable secure and flexible access across multiple network domains. It supports universal applicability by adjusting the types (e.g., adding PINs, passwords) and size of authentication sets to suit different scenarios. Encrypted authentication data can be integrated into smart cards or devices as needed. Data are obfuscated to ensure privacy and security, offering secure and flexible access in various scenarios.

(2) Our multi-factor authentication approach significantly reduces storage overhead compared to traditional attribute-based schemes \cite{11048743,huang2022blockchain}, where the CA must store each user’s full attribute and biometric sets. Instead, the CA maintains only the union of attributes for all users and a single verification token per user. For example, with 5 users each having 10 attributes and 10 biometric features, our approach stores only the union of these 50 attributes and 5 verification tokens, ranging from 15 to 55, while previous methods store all 100 factors individually. This efficiency scales in large networks, while also enhancing privacy, as attackers can access only the collective authentication set rather than individual user data.

(3) CDAuth improves cross-domain authentication efficiency by replacing costly bilinear pairings with lightweight native group operations, reducing terminal computation overhead. Performance analysis shows that, with the same number of attributes, the computational cost of the other two attribute-based cross-domain authentication schemes \cite{11048743,huang2022blockchain} is 3–6 times higher than that of CDAuth. Moreover, compared with multi-factor cross-domain authentication \cite{tian2024identity,9785665,shukla2022novel}, CDAuth incurs lower computational cost with the same number of authentication factors, making it both lightweight and efficient.

(4) We conducted a comprehensive security and performance analysis of CDAuth, evaluating its security and computational overhead. This detailed analysis ensures the scheme's robustness and efficiency.

\section{Related Work}
\label{sec:related work}
The widespread deployment of mobile terminals in the Internet of Things (IoT) has increased the demand for secure cross-domain access and interaction, making cross-domain authentication essential for trusted data exchange and reliable communication among multiple security domains. Recently, several blockchain-based cross-domain authentication schemes \cite{huang2022blockchain,GHAEMI2024103843} have been proposed to establish inter-domain trust via smart contracts and consensus mechanisms. Some schemes \cite{10489855,10509751} adopt blockchain for distributed cross-domain authentication, using decentralized key management to reduce management overhead and eliminate single points of failure. Blockchain’s decentralization, immutability, and traceability enhance authentication security, but often incurs high computational and communication costs. To improve efficiency, Jiang et al. \cite{jiang2024cross} combined Elliptic Curve Cryptography (ECC) and Certificateless Cryptography (CLC) in a two-layer consortium and private blockchain architecture, and Lv et al. \cite{lv2024reinventing} employed a multi-chain framework with proxy re-signatures to enable batch verification of multiple users using a single signature. Wang et al. \cite{10153598} combined edge computing with blockchain for cross-domain authentication, offloading pseudonym generation to edge servers to reduce latency and redundant interactions. Some methods \cite{10507750,9785665} used accumulators to improve efficiency, with blockchain maintaining dynamic accumulators per domain that accumulate key materials from terminals for batch authentication of unlinkable identities, enabling rapid anonymous authentication. Although these techniques reduce overhead to some extent, blockchain's inherent resource demands and slow read/write performance still limit their practical deployment.

Several non-blockchain solutions have been proposed that are both secure and efficient. Chen et al. \cite{article} introduced a certificate-based cross-domain authentication mechanism for VANETs, bidirectionally synchronized database to maintain authentication during primary database failures and mitigate single points of failure. It uses virtual identity anonymous authentication for privacy but faces efficiency and certificate management challenges. Additionally, \cite{PAN2023102821} used hyper-elliptic curve cryptography (HECC) to halve key length compared to ECC, and \cite{article2} replaced bilinear pairings with symmetric polynomials to reduce computational overhead, though further optimization is possible. Chen et al. \cite{article3} proposed a secure and efficient cross-domain authentication scheme integrating edge and fog computing, distributing computation and storage to the network edge for improved real-time interaction. It ensures forgery and collusion resistance and non-repudiation without blockchain. However, these schemes often rely on single-factor authentication, limiting scalability and resilience to attacks.

In IoT, multi-domain data interaction requires flexible authentication for dynamic access. Existing certificateless \cite{PAN2023102821} and identity/pseudonym-based methods \cite{article2} depend on identity keys, limiting flexibility and fine-grained access. The method in \cite{penuelas2024revocable} proposed an attribute-based authentication that supports flexible authentication with diverse attributes; however, it has seen limited adoption in cross-domain scenarios. He et al. \cite{11048743,huang2022blockchain} implemented an attribute-based cross-domain authentication scheme, but its computational cost remains prohibitively high. Zhang et al. \cite{9785665} proposed a multi-factor cross-domain authentication method that combines biometrics and hardware to generate key materials and employs accumulators for cross-domain verification. While this approach enhances security, its authentication granularity is coarse, scalability is limited, and hardware management and portability remain challenging—issues common to many existing multi-factor authentication schemes \cite{tian2024identity,shukla2022novel}.

Finally, Table \ref{tab11} provides a comparative overview of existing authentication schemes and  highlights their respective strengths and limitations. CDAuth optimizes these prior schemes by leveraging an edge–cloud collaborative framework to balance communication latency and enhance overall efficiency. It further refines multi-factor authentication to support broader applicability and fine-grained access control. Authentication factors are hidden, and credentials are self-generated, enabling bidirectional anonymous authentication via verification functions. Moreover, each communication round verifies data integrity and user status to ensure data immutability and timely detection of malicious behavior. These properties are achieved with reduced computational and storage overhead, making CDAuth suitable for complex network environments.

\begin{table*}[htbp]
\caption{Summary of related works}
\begin{center}
\renewcommand{\arraystretch}{2.7}
\scalebox{0.85}{
\begin{tabular}{ccccc}
\hline
Category & Scheme & Method & Advantage & Disadvantage \\
\hline
\multirow{12}{*}{\makecell{Blockchain-based}}&Jiang et al. \cite{jiang2024cross} (2024) & \makecell{Dual blockchain,\\ ECC, ID-based} & \makecell{Enhances security and privacy \\while reducing throughput} & \makecell{High blockchain read/write \\and maintenance complexity}\\
&He et al. \cite{11048743} (2025) & \makecell{Bilinear Pairing,\\Attribute-based} & \makecell{Batch verification\\Fine-grained and flexible} & \makecell{High blockchain read/write \\and maintenance complexity} \\
&Huang et al. \cite{huang2022blockchain} (2022) &\makecell{Bilinear Pairing,\\Zero-knowledge proof,\\Attribute-based} & \makecell{Anonymous authentication\\ Fine-grained and flexible} & \makecell{High blockchain read/write \\and maintenance complexity} \\
&Zhang et al. \cite{9785665} (2022) & \makecell{Dynamic Accumulators\\Multi-factor} & \makecell{Multi-factor anonymous authentication\\Enhances security and efficiency} & \makecell{High blockchain read/write \\and maintenance complexity}\\
&Cui et al. \cite{10384713} (2024) & \makecell{ECC, ID-based} & \makecell{Anonymous mutual authentication\\} & \makecell{High blockchain read/write \\and maintenance complexity} \\
&Ghaemi et al.\cite{GHAEMI2024103843} (2024) & \makecell{ECC, ID-based} & \makecell{Anonymous authentication\\More fault-tolerance} & \makecell{High blockchain read/write \\and maintenance complexity} \\
&Liu et al. \cite{10489855} (2024) & \makecell{ECC, ID-based} &  \makecell{Anonymous mutual authentication\\}& \makecell{High blockchain read/write \\and maintenance complexity} \\
&Sasikumar et al. \cite{10509751} (2024) & \makecell{ECC, ID-based} & Data transparency and traceability & \makecell{High blockchain read/write \\and maintenance complexity} \\
&Lv et al. \cite{lv2024reinventing} (2024) & \makecell{Bilinear Pairing,\\ Multi-blockchain,\\ID-based} & \makecell{Cross-chain authentication\\Batch verification} & \makecell{High blockchain read/write \\and maintenance complexity} \\
&Wang et al. \cite{10153598} (2024) & \makecell{ECC, ID-based} & \makecell{requiring only a single blockchain query\\low-latency edge response} & \makecell{High blockchain read/write \\and maintenance complexity}\\
&Zhong et al. \cite{10507750} (2024) & \makecell{ECC, ID-based} & \makecell{Batch authentication\\multi-accumulator batch proof} & \makecell{High blockchain read/write \\and maintenance complexity} \\
&Cheng et al. \cite{article2} (2023) & \makecell{ECC, ID-based} & \makecell{Pairing-free IBS\\Batch verification} & \makecell{High blockchain read/write \\and maintenance complexity}\\
\hline
\multirow{6}{*}{\makecell{Not blockchain-based}}&Chen et al. \cite{article} (2024) & \makecell{Bilinear Pairing, \\ID-based} & \makecell{Anonymous authentication\\ Batch verification} & \makecell{Relies on a trusted authority\\High computational overhead\\Coarse-grained and inflexible}\\
&Shukla et al. \cite{shukla2022novel} (2022) & \makecell{ECC,\\Multi-factor} & \makecell{Multi-factor anonymous authentication\\Enhances security} & \makecell{Risk of smart card loss\\Coarse-grained and inflexible} \\
&Tian et al. \cite{tian2024identity} (2024) & \makecell{Bilinear Pairing,\\Multi-factor} & \makecell{Multi-factor anonymous authentication\\Enhances security} & \makecell{Lacks cross-domain capability\\relies on a trusted authority\\Coarse-grained and inflexible}\\
&Pan et al. \cite{PAN2023102821} (2023) & \makecell{Bilinear pairing\\ID-based} & \makecell{Mutual authentication \\ Lightweight design} & \makecell{Lacks cross-domain capability\\Coarse-grained and inflexible} \\
&Penuelas et al. \cite{penuelas2024revocable} (2024) & \makecell{Bilinear Pairing,\\Attribute-based} & \makecell{Multi-authority,\\Fine-grained and flexible} & \makecell{High computational overhead\\Focuses on attribute-based encryption}\\
&Chen et al. \cite{article3} (2024) & \makecell{Bilinear pairing,\\Threshold signature} & \makecell{Distributed authentication\\Enhance security and robustness} & Complicated system deployment and upkeep\\

\hline
\end{tabular}
\label{tab11}
}
\end{center}
\vspace{-18pt}
\end{table*}

\section{Preliminaries}
We start by defining some notation used in this paper in Table \ref{tab123}. For integers $m$ and $n$ where $m\!\le\!n$, $[m,n]$ denotes the set $\{m,m+1,...,n \}$. If $m = 1$, we use $[n]$ to refer to $[1,n]$. For a prime $p$, $\mathbb{Z}_p$ denotes the set $\{0,1,...,p-1\}$ and $\mathbb{Z}_p^*$ denotes the set $[p-\!1] $, with arithmetic operations performed modulo $p$. $\lambda$ is a security parameter, and $p \ge 2^\lambda   + 1$. $||$ is the character connector.

\begin{table}[htbp]
\caption{Notations}
\vspace{-7pt}
\begin{center}
\scalebox{0.95}{
\begin{tabular}{ c p{5cm} }
\hline
\textbf{Notation} & \textbf{Description} \\
\hline
$\lambda$ & a security parameter \\
$D_m$ & the $m$-th subdomain \\
$u_{m,i}$ & the $i$-th user in the $m$-th subdomain \\
$id_{u_{m,i}}$, $pid_m$ & the pseudo-ID of $u_{m,i}$ and $ES_m$\\
$\mathcal{S}_T$ & the global attribute set in each subdomain \\ 
$\mathcal{S}$, $bio$ & each user’s attribute set, biometric template \\
$P\!K_{\!C\!S}$, $pk_{es_m}$, $pk_{u_{m,i}}$ & public keys of CS, $ES_m$ and $u_{m,i}$ \\
$ks_{m\!-\!M}$ & session key between $ES_m$ and $ES_M$ \\
$Aut_m$, $Aut_{m,i}$ & authentication credentials of $ES_m$, $u_{m,i}$ \\ 
$Aut_{M\!m,i}$ & cross-domain authentication credential in $D_M$ of $u_{m,i}$ \\
$reg_{u_{m,i}}$ & registration credential of $u_{m,i}$ \\
$Sig_m$, $Sig_{m,i}$ & Signatures of $ES_m$ and $u_{m,i}$ \\
$Sig_{M\!m,i}$ & Cross-domain signatures of $u_{m,i}$ for $D_M$ \\
\hline
\end{tabular}
}
\label{tab123}
\end{center}

\vspace{-23pt}
\end{table}

\subsection{Complexity assumption and Anti-Collision One-Way Hash Function}

\textbf{Definition 1. (Elliptic Curve Discrete Logarithm Problem, (DLP))} Let $\!(p,G,g)\!$ be a group system of order $p$. Given the tuple $\!(p,G,g,\zeta \cdot g)$, $g\!\in\!G$, the discrete logarithm problem states that a probabilistic polynomial-time (PPT) adversary $\mathcal{A}$ has negligible advantage in finding $\zeta\in \mathbb{Z}_p^*$. Formally, $\Pr [\mathcal{A}(p,G,g,\zeta \cdot g) = \zeta ]$ is negligible.

A one-way hash function $h$ is a deterministic mapping from an arbitrary-length binary string to a fixed-length $m$-bit output. Let the adversary 
$\mathcal{A}$’s advantage in finding a hash collision within time $rt$ be defined as $Adv_h^{\mathcal{A}}(rt)=\Pr[\mathcal{A}\to_r(x_1,x_2):x_1\neq x_2, h(x_1)=h(x_2)]$, where $\mathcal{A}\to_r(x_1,x_2)$ denotes two inputs selected randomly by $\mathcal{A}$. If $Adv_h^{\mathcal{A}}(rt)$ is negligible, then $h$ is collision-resistant.

\subsection{Fuzzy Extractor}
Biometric data, such as fingerprints, iris scans, or gaits, are inherently noisy and not perfectly deterministic. To securely extract keys from such data, we employ a fuzzy extractor \cite{Dodis2004}, consisting of two procedures: a probabilistic generation function $Gen$ and a deterministic reproduction function $Rep$.

(1) \textbf{Generation.} For a biometric template $Bio \!\in\!\! \mathcal{M}$, $Gen(Bio)\! = \!(\zeta, \rho)$, where $\zeta \!\!\in\!\! \{0,1\} ^\ell$ is an extracted string, $\rho \!\in\!\! \{ 0,1\} ^*$ is a public helper string and $\mathcal{M}$ is a metric space. If the statistic distance $SD((\zeta, \rho ),\!(U_\ell, \rho )) \!\le \!\varepsilon$, then $\zeta$ is indistinguishable from uniform even given $\rho$. $U_\ell$ is the uniform distribution over 
$\ell$-bit strings and 
$\varepsilon$ denotes the statistical bound on the deviation of $\zeta$ from uniform. Let $\mathcal{A}$ be a PPT adversary attempting to invert the fuzzy extractor generation function $Gen$. The advantage of $\mathcal{A}$ in recovering a biometric template $Bio$ from $(\zeta, \rho)\!=\!Gen(Bio)$ within time $rt$ is defined as $Adv_{Gen}^{\mathcal{A}}(rt)=\Pr[\mathcal{A}(1^\ell,\zeta, \rho)\to Bio': Gen(Bio')\! = \!(\zeta, \rho)]$, where $\mathcal{A}(1^\ell,\zeta, \rho)\to Bio'$ denotes that $\mathcal{A}$ outputs a candidate biometric template $Bio'$. If $Adv_{Gen}^{\mathcal{A}}(rt)$ is negligible in $\ell$, then $Gen$ is one-way.

(2) \textbf{Reproduction.} For a noisy biometric $Bio'$ with $dis(Bio,Bio' ) \le t$, $Rep(Bio',\rho)=\zeta$, where $dis(\cdot,\cdot )$ measures the distance between templates and $t$ is the error tolerance. Similarly, the adversary’s advantage in recovering $(Bio'',\rho')$ from $\zeta=Rep(Bio',\rho)$ is $Adv_{Rep}^{\mathcal{A}}(rt)=\Pr[\mathcal{A}(1^\ell,\zeta)\to (Bio'',\rho'): Rep(Bio'',\rho')\! = \!\zeta]$. If $Adv_{Rep}^{\mathcal{A}}(rt)$ is negligible in $\ell$, then $Rep$ is one-way.

\section{System Architecture and definitions} 
In this section, we introduce the system model, threat model, and the syntax definition of CDAuth.

\subsection{System model}
\begin{figure}[htbp]
\centerline{\includegraphics[width=9cm]{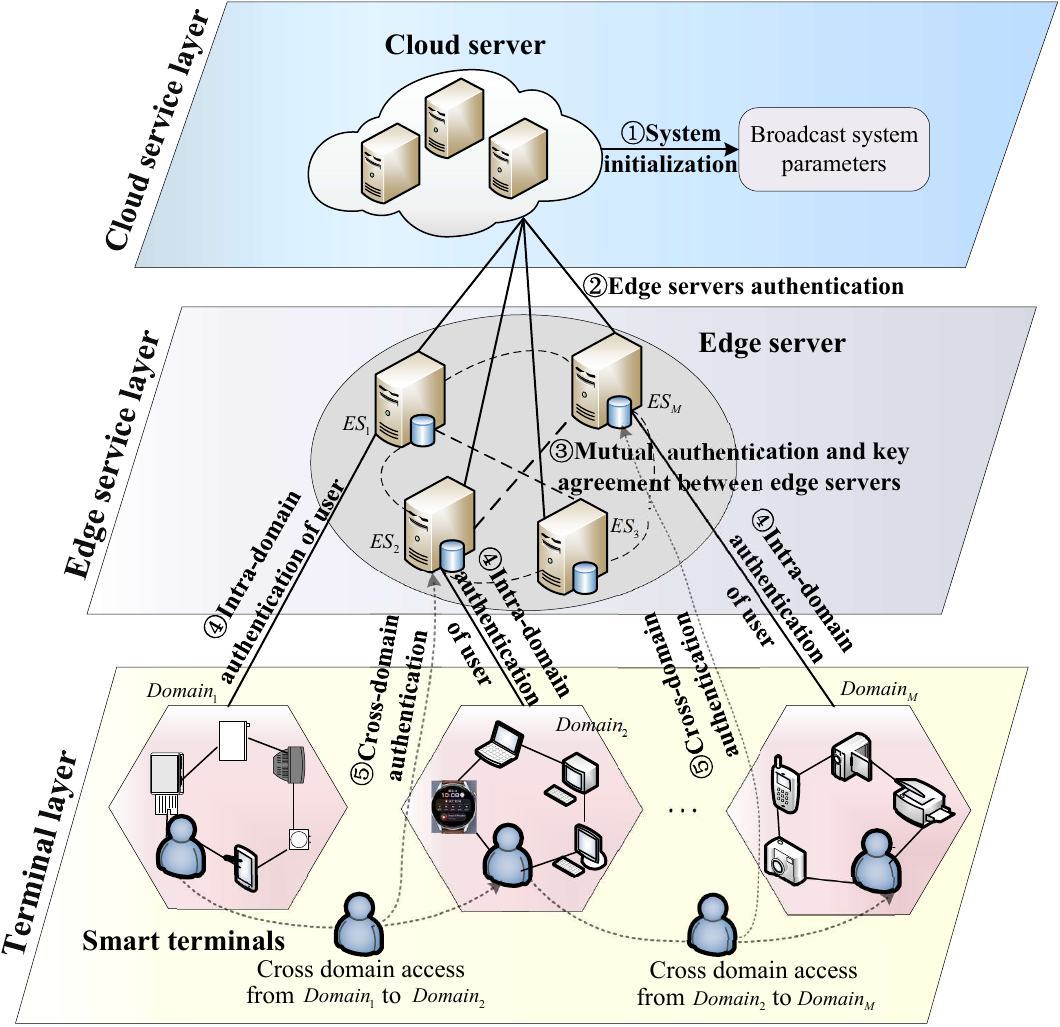}}
\caption{The system model}
\label{fig}
\vspace{-15pt}
\end{figure}

In this paper, a cross-domain multi-factor authentication protocol based on edge-cloud collaboration is designed. This system designs a three-layer model of cross-domain authentication: the cloud service layer, the edge service layer, and the terminal layer. As shown in Figure \ref{fig}, three layers work together to realize the intra-domain authentication and the cross-domain authentication of terminals.

The cloud service layer is composed of centrally deployed cloud servers (CSs) within large data centers, providing significant computing power and storage capacity. The CS, located far from terminal users, manages global data processing, storage, and complex tasks. Additionally, CS handles the initialization of edge servers (ESs) and terminal users, generating global system parameters, and overseeing the management and authentication of ESs. As a trusted entity, the CS acts as a CA for ESs, ensuring secure authentication. It also receives, analyzes, and processes data from ESs.

The edge service layer consists of edge servers with high computing power and storage. ESs, located near terminal users, manage real-time data processing to reduce cloud load and transmission delays. It handles terminal user intra-domain and cross-domain authentication, with ESs acting as CAs for terminal users. Additionally, this layer processes and analyzes data from terminal users (also called users) and enables data sharing in collaboration with the CS. 

The terminal layer includes mobile devices with limited computing power and storage. Terminals must be authenticated before network access. Terminal users obtain and use attribute and biometric ciphertexts for authentication with ESs, ensuring security and privacy. As terminal users operate independently and may collude, there is a risk of protocol attacks.

\subsection{Threat Model}
We assess system security using the widely adopted Dolev-Yao (DY) adversarial model \cite{dolev2003security}. The communication network has the following properties.

(1) When two entities communicate over a public channel, neither is fully trusted, leaving the exchange vulnerable to threats like man-in-the-middle, replay, or eavesdropping attacks. Each may try to access unauthorized information or modify data, compromising integrity and confidentiality.

(2) The adversary can be outside and inside. The adversary has excellent analytical and guessing abilities.

\subsection{The syntax definition of CDAuth} 
CDAuth is divided into five phases with twenty algorithms. These phases and algorithms are as follows.

(1) Initialization

$\bm{{\rm{{Setup}}}(\lambda)\!\! \to \!\!(pp,\!S\!K_{\!C\!S})}$: CS runs the setup algorithm, inputting the security parameter $\lambda$, to generate the public parameters $pp$ and the secret key $S\!K_{\!C\!S}$.

(2) Authentication of Edge Servers

$\bm{\textbf{ESSig}(pp)\!\!\to\!\! (Sig_m,\!pk_{es_m},\!sk_{es_m})}$: The signature algorithm is run by each $ES_m$. The algorithm takes the public parameters $pp$ as input and outputs the signature $Sig_m$ and the public-private key pair $pk_{es_m}, sk_{es_m}$ of $ES_m$.

$\bm{\textbf{VerESSig}(S\!K_{\!C\!S},\!pk_{es_m},\!Sig_m)\!\! \to \!\!(Aut_m,\!cf_m)}$: CS runs the signature verification algorithm that input $S\!K_{\!C\!S}$, $pk_{es_m}$,\! $Sig_m$ to produce the authentication credential factor $cf_m$ and the authentication credential $Aut_m$ for $ES_m$. 

$\bm{\textbf{ESAut}(pp,\! cf_m,\!sk_{es_m}\!) \!\!\to\!\! (Aut_m,\!\hbar _m\!)}$: Each $ES_m$ runs the authentication algorithm, using $cf_m$, $pp$, $sk_{es_m}$ to generate $Aut_m$ and the authentication credential validation factor $\hbar _m$.

$\bm{\textbf{VerESAut}(pp,\! \hbar _m\!)\!\! \to\!\! True/ \bot}$: CS runs the authentication verification algorithm, which takes $\hbar _m$ as input to generate either ``True" or ``$\bot$". If the result is ``True", CS broadcasts ``$ES_m$ authentication successful" in the system.

(3) Mutual Authentication and Key Agreement between Edge Servers

$\bm{\textbf{MulAut}(pp,\!Aut_m,\! pk_{es_M}\!)\!\!\to\!\! (cip_m,\!af_m\!)\!:}$ One party, $ES_m$, executes the mutual authentication algorithm, which takes $pp$, $Aut_m$, $pk_{es_M}$ as input and outputs the ciphertext $cip_m$ and the mutual authentication factor $af_m$. 

$\bm{\textbf{A\&KG}(pp,\!af_m,\!cip_m,\!pk_{es_m}\!)\!\!\! \to\!\!\! (\!cip_M,\!af_M,\!ks_{m\!-\!M}\!)\!\!:}$ The other party, $ES_M$, runs the authentication and session key generation algorithm, which inputs $pp$, $af_m$, $cip_m$, $pk_{es_m}$, authenticates $ES_m$ and generates the ciphertext $cip_M$, the mutual authentication factor $af_M$, and the session key $ks_{m-M}$. 

$\bm{\textbf{VA\&KG}(pp,\!af_M,\!cip_M,\!pk_{es_M}) \!\!\to \!\!(ks_{m\!-\!M})\!:}$ The authentication verification and session key generation algorithm is executed by $ES_m$, which takes $pp$, $af_M$, $cip_M$, $pk_{es_M}$ as inputs, verifies $ES_M$, and outputs the session key $ks_{m-M}$. 

(4) Intra-Domain Authentication of Terminals

$\textbf{Reg}\bm{(pp,\!au_{m,i},\!pk_{es_m}\!) \!\!\to \!\!(pk_{u_{m,i}},\!reg_{u_{m,i} },\!r\!f_{m,i}\!)\!:}$ $u_{m,i}$ runs the registration algorithm with input $pp$, $pk_{es_m}$, and the authentication information set $au_{m,i}$. It outputs the public/private key pair $pk_{u_{m,i}}$/$sk_{u_{m,i}}$, registration credential $reg_{u_{m,i} }$ and registration factor $rf_{m,i}$. 

$\textbf{VerReg}\bm{(pp,\!r\!f\!_{m,i},pk\!_{u\!_{m,i}},\!A\!U\!_m)\!\!\to\!\!\!(reg\!_{u\!_{m,i} }\!)}$: $ES_m$ runs the registration verification algorithm that intakes $pp$, $r\!f_{m,i}$, $pk_{u_{m,i}}$, and the global authentication information set $AU_m$. It outputs $reg_{u_{m,i} }$ and broadcasts ``$u_{m,i}$ registered successfully".

$\textbf{InitUAut}\bm{(pp,\!reg_{u_{m,i}},\!au_{m,i}^*,\!pk_{es_m }\!)\!\! \to\!\! (\!u\!f_{m,i}\!)\!\!:}$ $u_{m,i}$ runs the authentication initialization algorithm that input $pp$, $pk_{es_m}$, $reg_{m,i}$, $au_{m,i}^{*}$ to produce the authentication factor $u\!f_{m,i}$ for $ES_m$, where $au_{m,i}^{*}$ incorporates a perturbed biometric input $bio^*$ reflecting inherent measurement variability.

$\textbf{Iden\&AA}\bm{(pp,\!u\!f_{m,i},\!pk_{u_{m,i}}\!)\!\!\to\!\!(Aut_{m,i},\!ac\!f_{m,i}):}$ The user identification and authentication auxiliary algorithm is executed by $ES_m$. The algorithm takes $pp$, $u\!f_{m,i}$, $pk_{u_{m,i} }$ as input and outputs the authentication credential factor $acf_{m,i}$ and the authentication credential $Aut_{m,i}$.

$\textbf{UAut}\bm{(pp,\!ac\!f_{m,i},\!pk_{es_m}\!)\!\!\! \to\!\!\! (Aut_{m,i},\!V_{m,i}\!)\!\!:}$ Each $u_{m,i}$ runs the authentication algorithm, using $ac\!f_{m,i}$, $pp$, $pk_{es_m}$ to generate $Aut_{m,i}$ and the credential validation factor $V_{m,i}$.

$\textbf{VerUAut}\bm{(pp,\!V_{m,i},\!pk_{u_{m,i}}\!) \!\!\to\!\! True/\bot\!\!:}$ $ES_m$ runs the authentication verification algorithm, which takes $V_{m,i}$ as input to generate either ``True" or ``$\bot$". If the result is ``True", $ES_m$ broadcasts ``$u_{m,i}$ authentication successful".

(5) Cross-Domain Authentication of Terminals

$\textbf{CDReq}\bm{(pp,\!pid_m,u\!f_{\!m,i},\!Aut_{\!m,i},\!pk_{es\!_M}\!)\!\!\! \to \!\!\! (Req_{\!M})}$: The cross-domain authentication request algorithm is executed by $u_{m,i}$ with inputs $pp$, $pid_m$, $Aut_{m,i}$, $u\!f_{m,i}$, and $pk_{es_M}$, generating the request $Req_M$ for $ES_M$.

$\textbf{ESCom}\bm{(pp,\!Req_{\!M}\!,\!pk_{es_m}\!) \!\!\!\to\!\!\! (C\!om_{\!M})\!\!:}$ $ES_{\!M}$ executes the communication algorithm with inputs $pp$, $Req\!_M$, and $pk_{\!es_m}$, producing the communication request $Com_M$ for $ES_m$. 

$\textbf{MutESCom}\bm{(pp,C\!om_{\!M}\!,pk_{es_{\!M}}\!,pk_{u_{m,i}}\!) \!\!\!\to\!\!\! (C\!om_m\!)\!\!:}$ The mutual communication algorithm between $E\!S_m$ and $ES_{\!M}$ is executed by $E\!S_m$, takes $pp$, $C\!om_{\!M}$\!, $pk_{es_{\!M}}$\!, $pk_{u_{m\!,i} }$ as input, outputing the communication $C\!om_{\!m}$ for $E\!S_{\!M}$.

$\textbf{InCDAut}\bm{(pp,\!C\!om_m,\!pk_{u_{m\!,i}}\!)\!\!\! \to \!\!\!(Aut_{\!M\!m\!,i},\!C\!a\!f_{\!M\!m,i})\!\!:}$ $E\!S_{\!M}$ runs the cross-domain authentication initialization algorithm, which takes $pp$, $C\!om_m$, $pk_{u_{m\!,i}}$ as input and outputs the cross-domain authentication credential $\!Aut_{\!M\!m\!,i}$ and the cross-domain authentication credential factor $\!C\!a\!f_{\!M\!m,i}\!$ for $u_{m\!,i}$.

$\textbf{CDAut}\bm{(pp,\!C\!a\!f_{\!M\!m,i},pk_{es_{\!M}}\!)\!\!\! \to \!\!\!(Aut_{\!M\!m\!,i},\!V\!b_{m\!,i})\!\!:}$ $u_{m\!,i}$ runs the cross-domain authentication algorithm, using $\!C\!a\!f_{\!M\!m,i}$, $pp$, $pk_{es_{\!M}}$ to generate the cross-domain authentication credential $Aut_{\!M\!m\!,i}$ and the cross-domain authentication credential validation factor $\!V\!b_{m\!,i}$.

$\textbf{VerCDAut}\bm{(pp,\!V\!b_{m\!,i},pk_{u_{m\!,i}}\!) \!\!\to\!\! True/\bot\!\!:}$ $ES_M$ executes the cross-domain authentication verification algorithm on input $\!V\!b_{m,i}$, outputting either ``True” or ``$\bot$”. Upon a “True” result, $ES_M$ broadcasts ``$u_{\!m\!,i}$ cross-domain authentication successful” in $D_M$.

\section{Concrete construction
}
This section presents a detailed definition of the proposed CDAuth scheme. 

\subsection{Initialization}
Initialization is performed by the CS to generate the system parameters of the authentication system. The system is divided into $M$ subnet domains, each governed by an edge server $ES_m$, $m\in[M]$. 

$\bm{{\rm{{Setup}}}(\lambda)\!\! \to \!\!(pp,\!S\!K_{\!C\!S})}$:

\begin{enumerate}[label=\arabic*.]
\item Define $(p,G,g)$ as an additive cyclic group of prime order $p$, where $G$ is a group generated by $g$.

\item Randomly select $S\!K_{\!C\!S} \in \mathbb{Z}_p^*$ as the secret key and compute $P\!K_{\!C\!S}=S\!K_{\!C\!S}\cdot g$ as the public key.

\item Select two hash functions $H_1\!:\!\{ 0,1\} ^* \!\! \to \!\!\mathbb{Z}_p^* $, $H_2\!:\! G \!\! \to\!\! \mathbb{Z}_p^* $.

\item Randomly select $pid_m \in \mathbb{Z}_p^*$ for each $ES_m$ and add $ExpiryDate$ to avoid replay attacks.

\item Compute $T_{pid_m }  = t_{pid_m } g$, $w_m  = H_2 (pid_m ||T_{pid_m } )$ and $e_{pid_m }  = t_{pid_m }  + w_m S\!K_{\!C\!S}$.

\item Output the public parameters as

$pp = (p, G, g, P\!K_{\!C\!S}, H_1, H_2)$.
\end{enumerate}

CS publishes $pp$, which is made available to all system entities, while keeping the secret key 
$S\!K_{\!C\!S}$ confidential. CS secretly sends $(pid_m,T_{pid_m },e_{pid_m })$ to each $ES_m$.

\subsection{Authentication of edge servers}

\begin{figure}[htbp]
\centerline{\includegraphics[width=8cm]{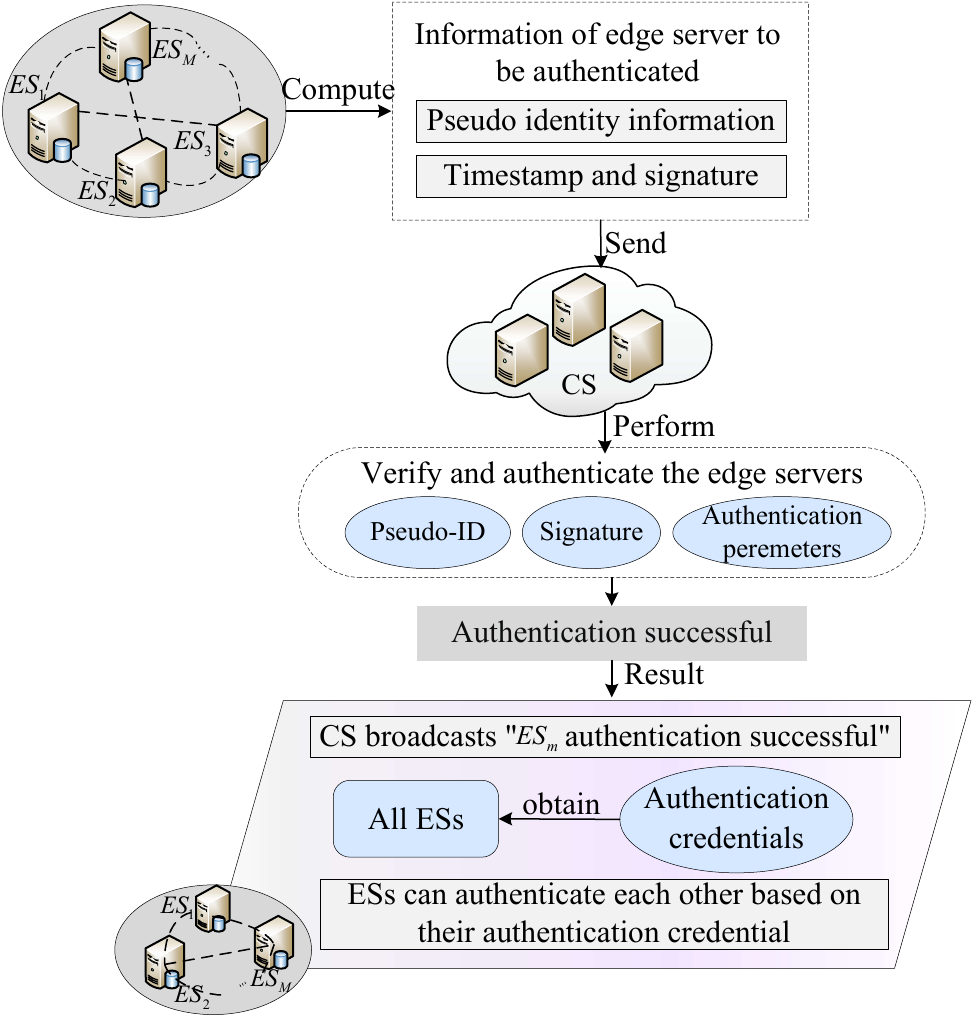}}
\caption{The edge server authentication process}
\label{fig1}
\vspace{-15pt}
\end{figure}

This section outlines the process of edge server authentication with the cloud server, as shown in Figure \ref{fig1}. 

$\bm{\textbf{ESSig}(pp)\!\!\to\!\! (Sig_m,pk_{es_m}\!,sk_{es_m})}$: After each $ES_m$ obtains $(pid_m, e_{pid_m }, T_{pid_m }) $, records the current timestamp $ts_m$ and performs the following steps.

\begin{enumerate}[label=\arabic*.]
\item Pick $sk_{es_m}  \in \mathbb{Z}_p^*$ as the secret key and computes $pk_{es_m}=sk_{es_m}g$ as the public key.

\item Pick $f_m  \in \mathbb{Z}_p^*$ and computes $F_m  = f_m g$, $\varpi _m  = H_2 (pid_m ||ts_m ||T_{pid_m } ||F_m )$, and $sig_m  = f_m  + e_{pid_m } \varpi _m$.

\item Output the signature as

$Sig_m = (pid_m ,ts_m ,F_m ,sig_m )$.
\end{enumerate}

$ES_m$ returns $Sig_m$ to CS, publishes $pk_{es_m}$, and keeps $sk_{es_m}$ secret.

$\bm{\textbf{VerESSig}(S\!K_{\!C\!S},pk_{es_m},\!Sig_m)\!\! \to \!\!(Aut_m,\!cf_m)}$: After CS obtains $Sig_m$, CS first checks the identity status of $ES_m$ and the correctness of $Sig_m$ by performing the following steps 1-3. Then, CS generates an authentication credential factor by steps 4-6.

\begin{enumerate}[label=\arabic*.]
\item Check $ExpiryDate$, if not hold, output $\bot$. Otherwise, $ES_m$ is in a normal state.

\item Compute $\varpi _m  = H_2 (pid_m ||ts_m ||T_{pid_m } ||F_m )$.

\item Check $sig_m  \cdot g = F_m  + \varpi _m (T_{pid_m }  + w_m P\!K_{\!C\!S} )$, if not hold, output $\bot$. Otherwise, $Sig_m$ is correct.

\item Compute $\eta _m \!\!=\!\!F_m \! S\!K_{\!C\!S}$, $\!\varphi _m \!\!=\!\!\eta _m pk_{es_m }\!S\!K_{\!C\!S}$.

\item Compute $Aut_m\!\!=\!\!H_2 (pid_m ||\eta _m )$, $V\!ut_m\!\!=\!\!H_2 (\eta _m )$.

\item Output the authentication credential factor as

$cf_m = (V\!ut_m ,\varphi _m  )$.

\end{enumerate}

CS keeps $Aut_m$ secret, and delivers $cf_m$ to $ES_m$.

$\bm{\textbf{ESAut}(pp,\! cf_m,\!sk_{es_m}\!) \!\!\to\!\! (Aut_m,\!\hbar _m\!)}$: After $ES_m$ obtains $cf_m$, $ES_m$ first checks the correctness of $cf_m$ by performing the following steps 1-2. Then, $ES_m$ generates an authentication credential by steps 3-5.

\begin{enumerate}[label=\arabic*.]
\item Compute $\eta _m  = \varphi _m /P\!K_{\!C\!S} sk_{es_m }$.

\item Check $H_2 (\eta _m )  = V\!ut_m$, if not hold, output $\bot$. Otherwise, $cf_m$ is correct.

\item Compute $Aut_m  = H_2 (pid_m ||\eta _m )$ as his authentication credential.

\item Compute $\hbar _m  = H_2 (pid_m ||T_{pid_m } )H_2 (\eta _m ||Aut_m)$ as the authentication credential validation factor. 

\item Output $Aut_m$, and $\hbar _m$.  

\end{enumerate}

$ES_m$ keeps $Aut_m$, and delivers $\hbar _m$ to CS.

$\bm{\textbf{VerESAut}(pp,\! \hbar _m\!)\!\!\! \to\!\! True/\bot}$: After CS obtains $\hbar _m$, it checks the correctness of $\hbar _m$ and $ES_m$'s authentication credential.

\begin{enumerate}[label=\arabic*.]
\item Compute $\mathchar'26\mkern-10mu\lambda _m  = w_m H_2 (\eta _m ||Aut_m )$.

\item Check $\mathchar'26\mkern-10mu\lambda _m = \hbar _m$, if not hold, output $\bot$. Otherwise, $ES_m$ is in a normal state, and authentication is complete. 

\item Output ``True" and bind the authentication credential $Aut_m$ to $ES_m$, then store it secretly.
\end{enumerate}

CS broadcasts ``$ES_m$ authentication successful" in the system.

\subsection{Mutual  authentication and key agreement between edge servers}

Upon successful authentication, $\{ES_m\}_{m\in[M]}$ collaborate in transaction processing to support cross-domain authentication. A mutual authentication and session key agreement scheme secures inter-server communication. Terminals can establish secure communication with others after completing intra- or cross-domain authentication in the following manner.

$\bm{\textbf{MulAut}(pp,\!Aut_m,\! pk_{es_M}\!)\!\!\to\!\! (cip_m,\!af_m\!)\!:}$

\begin{enumerate}[label=\arabic*.]
\item Compute $Au_m  = Aut_m sk_{es_m }^{-1}$.

\item Select $o_m ,temp_m  \in \mathbb{Z}_p^*$ by random, and compute $t_m  = o_m  + temp_m$.

\item Compute $cip_{m\!1}\!\!= \!\!o_msk_{\!es_m }\!pk_{\!es_{\!M} }\! $,\! $cip_{m\!2}\!\!=\!\!temp_msk_{\!es_{\!m} }\!{ pk_{\!es_{\!M} }\!}$, and set $cip_m  = (cip_{m1} ,cip_{m2} )$.

\item Compute $T_m  = H_2 (t_m g||Aut_mg ) sk_{es_m}pk_{es\!_M }$.

\item Output $cip_m$ and the mutual authentication factor as

$af_m = (t_m ,T_m, Au_m)$.
\end{enumerate}

$ES_m$ delivers $af_m$ and $cip_m$ to $ES_M$. 

$\bm{\textbf{A\&KG}(pp,\!a\!f_m,\!cip_m,\!pk_{es_m}\!)\!\!\! \to\!\! (\!cip\!_M,a\!f\!_M,ks_{m\!-\!M}\!)\!\!:}$ After $ES\!_M$ obtains $a\!f_m$ and $cip_m$, it first checks the identity of $ES_m$ and the correctness of $a\!f_m$, $cip_m$ by performing the following steps 1-2. $ES_M$ generates mutual authentication factors and a session key in steps 3–9 after successful verification.

\begin{enumerate}[label=\arabic*.]
\item Compute $p_{\!M\!m}  = t_m g$, $TT\!_M\!\!=\!\!H_2 (p_{\!M\!m}\! ||\!Au_m\!pk_{\!es_m\!}  )\!sk\!_{es\!_M}$.

\item Check $TT\!_Mpk_{es\!_m }\!\!=\!T_m$, if not hold, output $\bot$. Otherwise, $ES_m$ is legitimate, and $a\!f_m$, $\!cip_m$ are correct. 

\item Compute $Au_M  = Aut_M sk_{es_M }^{-1}$.

\item Get $o_m \!\! = \!\!sk_{es_M }^{ - 1}\! pk_{es_m }^{ - 1}\!cip_{m1}$, $temp_m \!\!=\!\! sk_{es_M }^{ - 1} \!pk_{es_A }^{ - 1}\!cip_{m2}$.

\item Randomly select $o_M  \in \mathbb{Z}_p^*$, compute $t_M  = o_M  + temp_m$.

\item Compute $cip_{\!M\!1} \!\!=\!\!sk_{\!es\!_M }\!pk_{\!es_m \!}o\!_M$, $cip_{\!M\!2}\!\!=\!\! sk_{\!e\!s\!_M }\!pk_{\!es\!_m }\! temp_m$, and set $cip\!_M  = (cip_{\!M\!1} ,\!cip_{\!M\!2} )$.

\item Compute $T\!_M  = H_2 (t\!_M g||Aut\!_Mg )sk_{es\!_M } pk_{es_m } $.

\item Generate the session key $ks_{m\!-\!M}  = H_2 (o_m ||o_M )$.

\item Output $cip_M$ and the mutual authentication factor as

$a\!f\!_M = (t_M ,TT_M, A\!Auu_M)$.
\end{enumerate}
$ES_M$ keeps $ks_{m\!-\!M}$ secret, and delivers $a\!f\!_M$, $cip\!_M$ to $ES_m$. 

$\bm{\textbf{VA\&KG}(pp,\!a\!f\!_M,\!cip\!_M,\!pk_{es\!_M}) \!\!\!\to\! \!\!(ks_{m\!-\!M})}$:  After $ES_m$ obtains $a\!f\!_M$ and $cip\!_M$, it first checks the identity of $ES\!_M$ and the correctness of $a\!f\!_M$, $cip\!_M$ following steps 1-2. $ES_m$ generates a session key in steps 3-4 after successful verification.

\begin{enumerate}[label=\arabic*.]
\item Compute $p_{\!m\!M}\!\! =\!t\!_M g$, $TT\!_m\!\! =\!\!H\!_2 (p_{\!m\!M} ||Au\!_Mpk_{es\!_M} )sk_{es\!_m }$.

\item Check $TT_mpk_{es\!_M } \!\!=\!\!T\!_M$, if not hold, output $\bot$. Otherwise, $ES\!_M$ is legitimate, and $\!a\!f\!_M$, $cip\!_M$ are correct. 

\item Get $o_M\!\!=\!\!sk_{es_m }^{ - 1}pk_{es_M }^{ - 1}\! cip_{M\!1}$, $temp_m\!\!=\!\!sk_{es_m }^{ - 1}pk_{es_M }^{ - 1}\! cip_{M\!2}$.

\item Output the session key as 

$ks_{m\!-\!M}  = H_2 (o_m ||o_M )$.
\end{enumerate}

$ES_m$ keeps $ks_{m\!-\!M}$ secret. 

\subsection{Intra-Domain Authentication for Terminals}
\label{123}
\begin{figure}[htbp]
\centerline{\includegraphics[width=8cm]{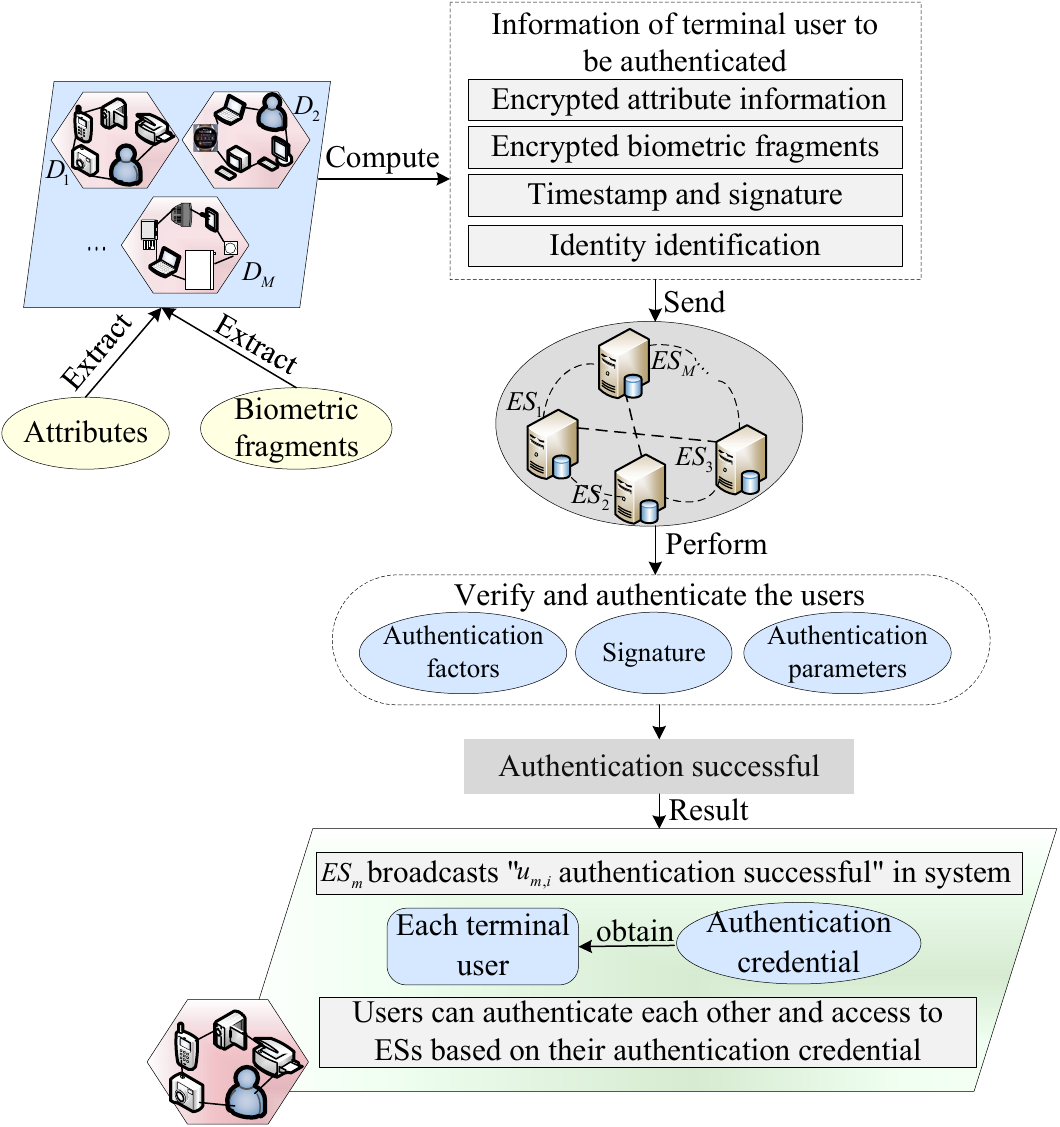}}
\caption{The intra-domain authentication process}
\label{fig3}
\vspace{-15pt}
\end{figure}

This section presents the intra-domain authentication process between terminals and edge servers (Figure \ref{fig3}). Let $u_{m,i}$ denote the $i$-th user in subdomain $D_m$, identified by a pseudo-ID $id_{u_{m,i}}$ embedding an expiration timestamp $ExpiryDate$ to resist replay attacks. In each subdomain, the edge server $ES_m$ retrieves each user’s attribute set $\mathcal{S}$ as a subset of the global attribute set $\mathcal{S}_T = \{A_1, A_2, \dots, A_R\}$, i.e., $\mathcal{S} = \{A_j\}_{j \in r}$ with $r\subseteq [R]$. It also obtains each user’s biometric template $bio=\{Bio_1,Bio_2,\dots,Bio_k\}$, which may include fingerprints, iris patterns, facial structures, and gaits \cite{wei2022towards,meng2024bioinspired}. For each user, $ES_m$ computes $Gen(bio)=(\zeta_{m,i},\rho_{m,i})$ and derives the verification token $VT_{m,i}=H_1(A_j||H_1(\zeta_{m,i}))$. By aggregating all users’ tokens, it constructs a global verification set $\mathcal{V}_T=\{VT_{m,i}\}_{i\in[n]}$, where $n$ denotes the number of users in subdomain $m$. Consequently, $ES_m$ stores the global authentication information $AU_m=\mathcal{S}_T\cup \mathcal{V}_T$, while each user holds a personalized authentication set $au_{m,i}=\mathcal{S}\cup bio$.

$\textbf{Reg}\bm{(pp,\!au_{m,i},\!pk_{es_m}\!) \!\!\to \!\!(pk_{u_{m,i}},\!reg_{u_{m,i} },\!r\!f_{m,i}\!)\!:}$

\begin{enumerate}[label=\arabic*.]
\item Randomly select $sk_{u_{m,i}}\!\! \in \!\!\mathbb{Z}_p^* $ as the serect key and compute the public key $pk_{u_{m,i}}  = sk_{u_{m,i}} g$.

\item For each attribute $A_j \in au_{m,i}$, compute $H_1(A_j)$.

\item Extract $bio$ from $au_{m,i}$, compute $Gen(bio) = (\zeta _{m,i} ,\rho _{m,i} )$, $BIo_{m,i}  = H_1 (\zeta _{m,i} )$.

\item Compute $V\!t_{m,i}=H_1 (H_1 (A_j)||BIo_{m,i})$.

\item Compute $reg_{u_{m,i} }\!\!=\!H_1(\sum\nolimits_{j \in r}\!H_1 (A_j)\!+\!BIo_{m,i})$.

\item Record the current timestamp $ts_{u_{m,i}}$ and compute $U\!V\!e_{m,i}  = id_{u_{m\!,i} } ts_{u_{m\!,i}} V\!t_{m\!,i}sk_{u_{m\!,i} } pk_{es_m }$.

\item Output $reg_{u_{m,i} }$ and the registration factor as 

$\!r\!f_{\!m,i}\!\!= (id_{u_{m,i}},\!ts_{u_{m,i}},\{H_1\!(A_j)\}_{j\in r},B\!I\!o_{\!m,i},\!U\!V\!e_{\!m,i})$.
\end{enumerate}

Then, $u_{m,i}$ publishes $pk_{u_{m,i}}$, keeps $sk_{u_{m,i}}$, $reg_{u_{m,i} }$ secret, and delivers $\!r\!f_{\!m\!,i}$ to $ES_m$.

$\textbf{VerReg}\bm{(pp,\!r\!f_{\!m\!,i},pk_{u_{m\!,i}},\!AU\!_m)\!\!\to\!\!\!(reg_{u_{m,i} }\!)\!\!:}$ After $ES_m$ obtains $\!r\!f_{\!m\!,i}$, it first checks the status of $u_{\!m\!,i}$ and the correctness of $\!r\!f_{\!m\!,i}$ by performing the following steps 1-2. Then, $ES_m$ generates a registration credential in steps 3-6.
\begin{enumerate}[label=\arabic*.]
\item Compute $V\!e_{m\!,i}=H_1 (H_1 (A_j)||BIo_{m,i}))id_{\!u_{\!m\!,i} } ts_{\!u_{\!m\!,i}}$.

\item Check $\!V\!e_{\!m\!,i}sk_{es\!_m }pk_{\!u_{\!m\!,i} }\!\! = \!\!U\!V\!e_{\!m\!,i}$, if not hold, output $\bot$. Otherwise, $u_{m\!,i}$ is in a normal state, and $\!r\!f_{\!m\!,i}$ is correct. 

\item For each attribute $A_j \in AU_m$, compute $H_1\!(A_j )$.

\item Compute $\{H_1\!(\!A_j\! )\}_{j\in [R]}\!\! \cap\!\! \{H_1 (A_j)\}_{j \in r}$ to determine $\{A_j\}_{j\in r}$ and $VT_{m,i}'=H_1(A_j||BIo_{m,i})$. 

\item Check     $V\!T_{\!m\!,i}'\!\!=\!\!\!V\!T_{\!m\!,i}$, if not hold, output $\bot$ and remove $u\!_{m\!,i}$ from the network. Otherwise, $u_{\!m\!,\!i}$ registers successfully.

\item Output the registration credential $reg_{u_{m,i} }$ as 

$reg_{u_{m\!,i} }\!\!=\!\!H_1(\sum\nolimits_{j \in r}\!H_1 (A_j )\!+\!B\!I\!o_{m\!,i}')$.
\end{enumerate}

Finally, $ES_m$ keeps $reg_{u_{m\!,i} }$ secret, and broadcasts ``$u_{m\!,i}$ registered successfully" in domain $D_m$.

$\textbf{InitUAut}\bm{(pp,\!reg_{u_{m,i}},\!au_{m,i},\!pk_{es_m }\!)\!\! \to\!\! (u\!f_{m\!,i})\!\!:}$

\begin{enumerate}[label=\arabic*.]

\item Pick $b_j,c \in \mathbb{Z}_p^*$ randomly and compute $R_{au_{A,j} }  = b_j A_j g$, $R_{au_{B} }  = c \zeta_{m,i} g$ to realize anonymity.

\item Get $dd_{\!u_{m\!,i} }\!\! = \!reg_{u_{\!m\!,i} }\!H_2 \!(R_{au_{\!A\!,j} }\!||\!R_{\!au_{\!B} } ) sk_{\!u_{m\!,i} }$.

\item Output the authentication factor

$u\!f_{m,i}  = (id_{u_{m,i}},dd_{u_{m,i} },\{R_{au_{\!A,j} }\}_{j\in r},R_{au_{\!B} })$.
\end{enumerate}

$u_{m,i}$ sends $u\!f_{m,i}$ to $ES_m$.

$\textbf{Iden\&AA}\bm{(pp,\!u\!f_{\!m\!,i},\!pk_{u_{m\!,i}}\!)\!\!\to\!\!(Aut_{m\!,i},\!acf_{\!m\!,i}):}$ After $ES_m$ obtains $u\!f_{m\!,i}$, it first checks the identity status of $u_{m,i}$ and the correctness of $u\!f_{m\!,i}$ following steps 1-2. Then, $ES_m$ generates an authentication credential factor in steps 3-6. 

\begin{enumerate}[label=\arabic*.]
\item Compute $d_{u_{\!m\!,i} }\!\! =\! \!H_2\!(R_{\!au_{\!A\!,j} }\!||\!R_{\!au_{\!B} } )reg_{\!u_{\!m\!,i} }pk_{u_{m,\!i} }$.

\item Check $d_{u_{\!m\!,i} } \!\!= \!\!dd_{\!u_{\!m\!,i}}g$, if not hold, output $\bot$. Otherwise, $u_{m\!,i}$ is in a normal state, and $u\!f_{\!m\!,i}$ is correct. 

\item Randomly select $\beta\!_j,\!\delta \!\!\in\!\mathbb{Z}_p^*$ and compute $R_{au_{A,\!j} }' \!\!\!=\!\! \beta _j\!R_{au_{A,\!j} }$, $R_{au_{B} }' \!\! =\! \delta R_{au_{B} }$.


\item Compute $Sig_{m,i}\!\!=\!pid_msk_{es_m }H_2 (\beta _j\! A_jg||\delta \zeta _{m,i} g)$.

\item Compute $Aut_{m,i} = \sum\nolimits_{j \in r}\beta _j A_j g + \delta \zeta _{m,i} g$.

\item Output $Aut_{m,i}$ and the authentication credential factor as 

$acf_{m,i} = (pid_m, Sig_{m,i}, \{R_{au_{A,j} }'\}_{j\in r}, R_{au_{B} }')$.
\end{enumerate}

$ES_m$ keeps $Aut_{m,i}$ secret, and sends $acf_{m,i}$ to $u_{m,i}$.

$\textbf{UAut}\bm{(pp,\!acf_{m,i},\!pk_{es_m}\!)\!\!\! \to\!\!\! (Aut_{m,i},\!V_{m,i}\!)\!\!:}$ After $u_{m,i}$ obtains $acf_{m,i}$, $u_{m,i}$ first checks the identity status of $ES_m$ and the correctness of $acf_{m,i}$ following steps 1-2. Then, $u_{m,i}$ generates the authentication credential in steps 3-4. 

\begin{enumerate}[label=\arabic*.]
\item Compute $RR_{au_{A,j} }  = b_j^{ - 1} R_{au_{A,j} }' $, $RR_{au_{B} } = c^{ - 1} R_{au_{B} }'$.

\item Check $pid_m\!H_2\! (RR_{au_{A\!,j} }\!||\!RR_{au_{B} } )pk_{es_m}\!\!=\!\!Sig_{m\!,i}g$, if not hold, output $\bot$. Otherwise, $ES_m$ is in a normal state, and $acf_{m,i}$ is correct. 

\item Compute $Aut_{m,i}  = \sum\nolimits_{j \in r}RR_{au_{A,j}} + RR_{au_{B}}$ as the authentication credential.

\item Compute $V_{m,i}  = H_2 (Aut_{m,i} )sk_{u_{m,i} }^{ - 1}$.

\end{enumerate}

$u_{m,i}$ keeps $Aut_{m,i}$ secret, and delivers $id_{u_{m,i}}$, $V_{m,i}$ to $ES_m$.

$\textbf{VerUAut}\bm{(pp,\!V_{\!m\!,i},\!pk_{\!u_{m\!,i}}\!) \!\!\to\!\! True/\bot\!\!:}$ After $E\!S_m$ obtains $id_{\!u_{\!m\!,i}}$ and $V_{\!m\!,i}$, $E\!S_m$ checks the identity status and the authentication credential of $u_{\!m\!,i}$ by performing the following steps. 

\begin{enumerate}[label=\arabic*.]
\item Check $V_{\!m\!,i}pk_{\!u_{\!m\!,i}} \!\!=\!\! H_2\!(Aut_{m\!,i} )g$, if not hold, output $\bot$. Otherwise, $u_{m\!,i}$ is in a normal state, and $u_{m\!,i}$ has completed authentication.

\item Output ``True" and bind the authentication credential $Aut_{m,i}$ to $u_{m,i}$, and store it secretly.
\end{enumerate}

$ES_m$ broadcasts ``$u_{m,i}$ authentication successful" in $D_m$.
 
\subsection{Cross-domain Authentication of Terminals}
\label{15}
\begin{figure}[htbp]
\centerline{\includegraphics[width=9cm]{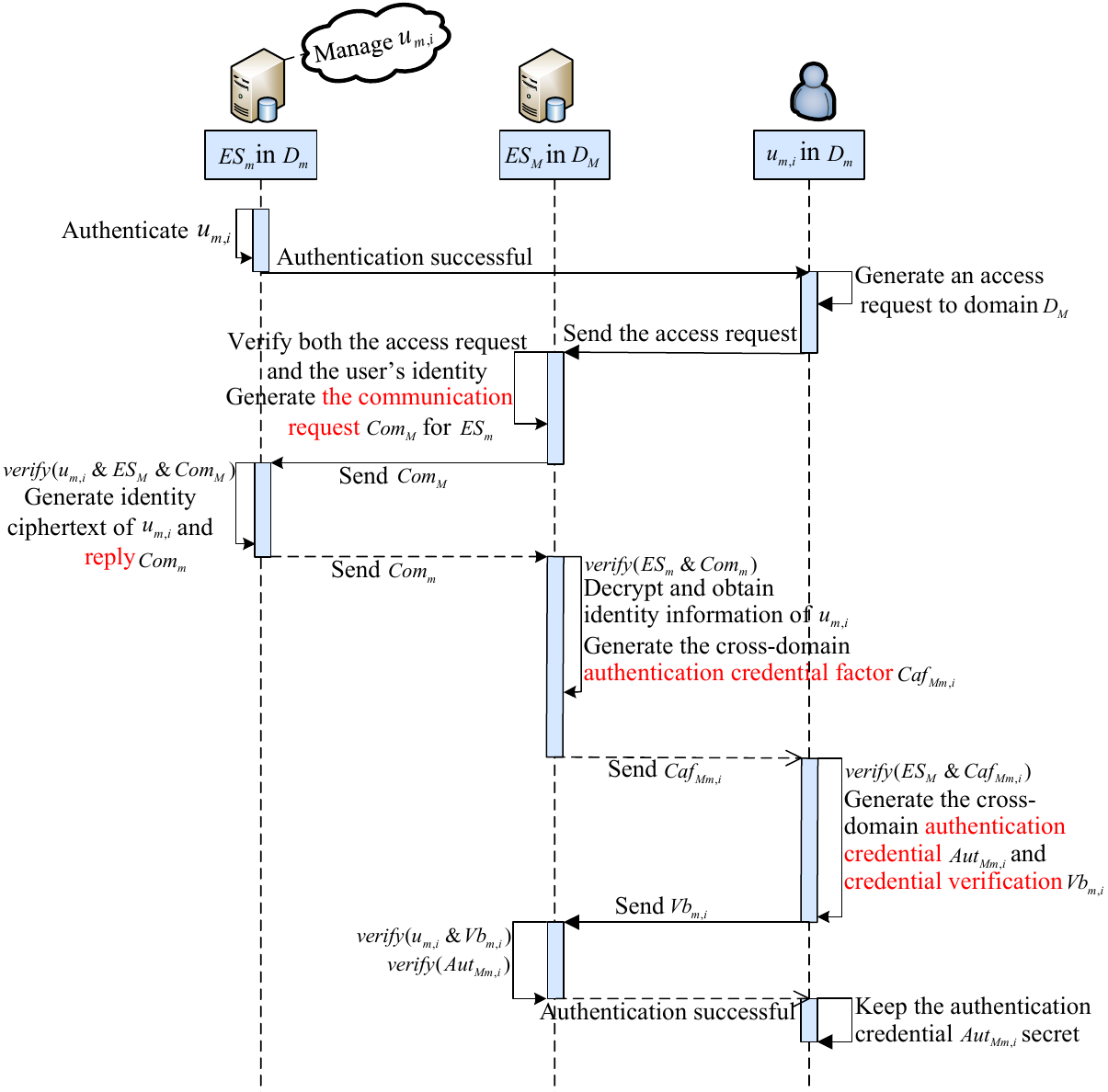}}
\caption{The cross-domain authentication process}
\label{fig4}
\vspace{-15pt}
\end{figure}

Cross-domain authentication allows terminal users to securely access resources in other sub-domains. The process whereby an authenticated user $u_{m,i}$ from domain $D_m$ requests authentication from the edge server $ES_M$ in domain $D_M$ is detailed below and illustrated in Figure \ref{fig4}.

$\textbf{CDReq}\bm{(pp,\!pid_m,\!u\!f_{m,i},Aut_{m,i},\!pk_{es\!_M}\!)\!\! \to \!\! (Req_M)\!:}$

\begin{enumerate}[label=\arabic*.]
\item Compute $V\!A_{m\!,i} \!\!=\!\! H_2(Aut_{m\!,i})\!H_2 (R_{au_{A\!,j} }||R_{au_{B} })pid_m sk_{u_{m,i}}^{-1}$ and  $AAu_{u_{m,i}}  = H_2(Aut_{m,i})pk_{es\!_M}$.

\item Output the access request for $D_M$ as

$Req_M\!\!=\!\!(pid_m,\!id_{\!u_{\!m\!,i}},V\!A_{\!m\!,i},A\!A\!u_{\!u_{\!m\!,i}},R_{\!au_{\!A\!,j} },R_{\!au_{\!B} })$.
\end{enumerate}

$u_{m,i}$ sends $Req_M$ to $\!E\!S\!_M$.

$\textbf{ESCom}\bm{(pp,\!Req\!_M\!,pk_{\!es\!_m}\!) \!\!\!\to\!\!\! (C\!om\!_M)\!\!:}$ After $\!E\!S\!_M$ obtains $Req_M$, it first checks the identity status of $u_{\!m\!,i}$ and the correctness of $Req_M$ following steps 1-2. Then, $ES\!_M$ communicates with $ES_m$ to verify $u_{m,i}$ in steps 3-5. 

\begin{enumerate}[label=\arabic*.]
\item Compute $R\!Au_{\!m\!,i}\!\! =\! pid_m\!H_2 (\!R_{\!au_{A\!,j} }\!||\!R_{au_{\!B} })$.

\item Check $R\!Au_{\!m\!,i}AAu_{u_{\!m\!,i}}sk_{es_M}^{-1}\!\!=\!\!V\!A_{m\!,i}pk_{u_{\!m\!,i}}$, if not hold, output $\bot$. Otherwise, $ES_m$ is in a normal state, and $Req\!_M$ is correct.

\item Compute $\theta _M  = H_2(Aut_M) H_1 (pid_M||id_{u_{m,i}} ) pk_{es_m }$.

\item Compute $AAu_M  = H_2(Aut_M) sk_{es_M }^{-1}$.

\item Output the communication request to $ES_m$ as

$Com\!_M\!\!=\!\!(id_{u_{m\!,i} },V\!A_{m\!,i},pid_M,\! A\!A\!u_M,\!\theta _M)$.
\end{enumerate}
$\!E\!S\!_M$ sends $C\!om\!_M$ to $ES_m$.

$\textbf{MutESCom}\bm{(pp,\!C\!om\!_M\!,pk_{es\!_M}\!,pk_{u_{m\!,i}}\!) \!\!\!\to\!\!\! (C\!om_m\!)}$: Upon receiving $C\!om\!_M$, $ES_m$ verifies the identity status of $ES_M$, $u_{m,i}$ and the validity of $Com_M$ through steps 1–4, and transmits the identity ciphertext of $u_{m,i}$ to $ES_M$ in steps 5–7.

\begin{enumerate}[label=\arabic*.]
\item Compute $H\!A_M=AAu_M pk_{es_M }$.

\item Check $H\!A_MH_1 (pid_M||id_{u_{m,i}} )  = \theta _M sk_{es_m}^{-1}$, if not hold, output $\bot$. Otherwise, $ES_m$ is in a normal state.

\item Compute $R\!Au_{m\!,i}'\!=\!pid_m\!H_2 (R_{\!au_{A\!,j} }\!||\!R_{au_{\!B} }  )$.

\item Check $H\!A_M\!R\!Au_{m\!,i}' \!\!= \!\!pk_{u_{m\!,i} }\! V\!A_{m\!,i}$, if not hold, output $\bot$. Otherwise, $u_{m,i}$ is in a normal state, and $Com_M$ is correct.

\item Encrypt $C_j \!=\! ks_{m\! -\! M}\! A_jg$ and $C_B\!  = \!ks_{m\!-\!M}\zeta_{m,i}g$ by using $ks_{m\!-\!M}$ and sets $C_{u_{m,i} }  = (C_j ,C_B )$.

\item Compute $\theta _m  = H_2(Aut_m)H_1 (pid_m||id_{u_{m,i}} ) pk_{es_M }$ and $AAu_m  = H_2(Aut_m) sk_{es_m }^{-1}$.

\item Output the identity ciphertext $C_{u_{m,i} }$ and set

$Com_m=(pid_m, \theta _m,A\!A\!u_m, C_{\!u_{m,i} })$.

\end{enumerate}

$ES_m$ sends $C\!om_m$ to $E\!S\!_M$.

$\textbf{InCDAut}\bm{(pp,\!C\!om_m,\!pk_{u_{m,i}}\!)\!\!\! \to \!\!\!(\!Aut_{\!M\!m\!,i},\!C\!a\!f_{\!M\!m\!,i}\!)\!\!:}$ After receiving $C\!om_m$, $\!E\!S\!_M$ checks the identity status of $ES_m$ and the correctness of $C\!om_m$ following steps 1-2. Then, $\!E\!S\!_M$ generates the authentication credential factor in steps 3-7. 

\begin{enumerate}[label=\arabic*.]
\item Compute $H\!A_m  = A\!A\!u_m pk_{es_m }$, $H\!a_m  = \theta _msk_{\!es\!_M }^{-1}$.

\item Check $H\!A_m\!H_1\!(pid_m||id_{u_{m,i}} )\!\! =\!\!H\!a_m$, if not hold, output $\bot$. Otherwise, $E\!S\!_m$ is in a normal state, $C\!om_m$ is correct. 

\item Decrypt $C_{u_{m,i} }$ to obtain $A_jg$ and $\zeta _{m,i}g$ as $A_jg = C_jks_{m-M}^{-1}$, $\zeta _{m,i}g = C_Bks_{m-M}^{-1}$.

\item Pick $\beta _j',\!\delta'\!\! \in \!\!\mathbb{Z}_p^*$ and compute $R_{au_{\!A\!,j} }^{''}\!\!=\!\!\beta _j' R_{au_{A\!,j} }$,$R_{au_{B} }^{''} =\delta' R_{au_{B} }$.

\item Compute $Sig_{\!M\!m,i}\!\!  = \!\!pid_{\!M}sk_{es_{\!M} } pk_{\!u_{m\!,i} }H_2(\beta _j' A_j g||\delta' \zeta _{m,i}g ) $.

\item Compute $Aut\!_{M\!m,i} = \sum\nolimits_{j \in r}\beta _j' A_j g + \delta' \zeta _{m,i} g$.

\item Output $Aut_{\!M\!m\!,i}$ and the authentication credential factor $C\!a\!f_{\!M\!m\!,i}$ as

$C\!a\!f_{\!M\!m\!,i} = (pid\!_M, Sig_{\!M\!m\!,i}, \{R_{au_{\!A\!,j} }''\}_{j\in r}, R_{au_{\!B} }'')$.
\end{enumerate}

$\!E\!S\!_M$ keeps $Aut_{\!M\!m\!,i}$ secret, and sends $\!C\!a\!f_{\!M\!m\!,i}$ to $u_{m,i}$.

$\textbf{CDAut}\bm{(pp,\!C\!a\!f_{\!M\!m\!,i},pk_{\!es\!_M})\!\! \to \!\!(Aut_{\!M\!m\!,i},\!V\!b_{m\!,i})\!\!:}$ After receiving $\!C\!a\!f_{\!M\!m\!,i}$, $u_{\!m\!,i}$ checks the identity status of $\!E\!S\!_M$ and the correctness of $\!C\!a\!f_{\!M\!m\!,i}$ following steps 1-2. Then, $u_{\!m\!,i}$ generates the authentication credential in steps 3-4. 

\begin{enumerate}[label=\arabic*.]
\item Compute $R\!R\!B_{au_{A\!,j} }\!\! = \!b_j^{ - 1}\!R_{au_{A\!,j} }''$, $R\!R\!B_{au_{B} } \!\!=\! c^{ - 1}\!R_{au_{B} }''$.

\item Check $pid_{\!M}\!sk_{u_{m\!,i}}pk_{es\!_M}\!H_2\!(R\!R\!B_{\!au_{\!A\!,j} }\!||\!R\!R\!B_{\!au_{\!B} } )\!\! =\!\!Sig_{\!M\!m\!,i}$, if not hold, output $\bot$. Otherwise, $\!E\!S\!_M$ is in a normal state, and $\!C\!a\!f_{\!M\!m\!,i}$ is correct. 

\item Compute $Aut_{\!M\!m\!,i} \!\!=\!\! \sum\nolimits_{j \in r}\!R\!R\!B_{\!au_{\!A\!,j}}\!+\!R\!R\!B_{\!au_{\!B}}$ as the authentication credential.

\item Output $V\!b_{m,i} \!=\! H_1 (id_{u_{m,i} } )\!H_2 (Aut_{\!M\!m,i} )sk_{u_{m,i} }^{ - 1}$.

\end{enumerate}

$u_{m,i}$ keeps $Aut_{\!M\!m\!,i}$ secret, delivers $id_{u_{m,i}}$, $V\!b_{m,i}$ to $E\!S\!_M$. 

$\textbf{VerCDAut}\bm{(pp,\!V\!b_{\!m\!,i},\!pk_{u_{m\!,i}}\!) \!\!\to\!\! True/\bot\!\!:}$ After $\!E\!S\!_M$ obtains $id_{u_{m\!,i}}$, $\!V\!b_{\!m\!,i}$, $\!E\!S\!_M$ checks the identity status and the authentication credential of $u_{m\!,i}$ using the following steps. 

\begin{enumerate}[label=\arabic*.]
\item Check $V\!b_{\!m\!,i}pk_{\!u_{\!m\!,i}}\!\!=\!\! H_2\!(Aut_{\!M\!m\!,i} )\!H_1 (id_{\!u_{\!m\!,i}} )g$, if not hold, output $\bot$. Otherwise, $u_{m\!,i}$ is in a normal state, and $u_{m\!,i}$ has completed cross-domain authentication.

\item Output ``True" and bind the authentication credential $Aut_{\!M\!m\!,i}$ for $\!D\!_M$ to $u_{\!m\!,i}$ and secretly store.
\end{enumerate}

$\!E\!S\!_M$ broadcasts ``$u_{\!m\!,i}$ cross-domain authentication successful" in $\!D\!_M$.

\section{Correctness and security analysis}
\label{1}
This section presents the correctness and the security analysis of the proposed scheme.

\subsection{Correctness}
This section demonstrates the correctness of CDAuth, showing that all entities complete the authentication as intended.

(1) In the edge server authentication phase, $ES_m$ generates its authentication credential $Aut_m$ using the $\textbf{ESAut}$ algorithm and transmits the corresponding validation factor $\hbar_m$ to the CS. Upon receiving $\hbar_m$, the CS verifies the following relation
\vspace{-3pt}
\begin{equation*}
\begin{split}
\mathchar'26\mkern-10mu\lambda _m &=w_m H_2 (\eta _m ||Aut_m ) \\&=H_2 (pid_m ||T_{pid_m } )H_2 (\eta _m ||Aut_m )\\&=\hbar _m.
\end{split}
\end{equation*}

Since both sides of the equation are equal, the correctness of $Aut_m$ is guaranteed, and the authentication process succeeds.

(2) In intra-domain authentication, the terminal user $u_{m,i}$ executes the $\textbf{UAut}$ algorithm to produce an authentication credential $Aut_{m,i}$, and sends the corresponding validation factor $V_{m,i}$ to $ES_m$. After receiving $V_{m,i}$, $ES_m$ verifies
\vspace{-3pt}
\begin{equation*}
\begin{split}
V_{\!m\!,i}pk_{\!u_{\!m\!,i}} \!\!&=\!\! H_2 (Aut_{m,i} )sk_{u_{m,i} }^{ - 1}pk_{\!u_{\!m\!,i}}\\&=H_2 (Aut_{m,i} )g.
\end{split}
\end{equation*}

The satisfaction of the equation confirms the correctness of $Aut_{m,i}$, thereby completing the authentication successfully.

(3) In cross-domain authentication, $u_{m,i}$ runs the $\textbf{CDAut}$ algorithm to produce a cross-domain authentication credential $Aut_{Mm,i}$ for $D_M$, and sends the corresponding validation factor $Vb_{m,i}$ to $ES_M$. After receiving $Vb_{m,i}$, $ES_M$ verifies
\vspace{-3pt}
\begin{equation*}
\begin{split}
V\!b_{\!m\!,i}pk_{\!u_{\!m\!,i}}\!\!&=\!\!H_1 (id_{u_{m,i} } )\!H_2 (Aut_{\!M\!m,i} )sk_{u_{m,i} }^{ - 1}pk_{\!u_{\!m\!,i}} \\&=H_2\!(Aut_{\!M\!m\!,i} )\!H_1 (id_{\!u_{\!m\!,i}} )g.
\end{split}
\end{equation*}

As the equality holds, the validity of $Aut_{\!M\!m,i}$ is established, ensuring the success of the cross-domain authentication.

\subsection{Security analysis}
This section presents both the formal and informal security analysis of the proposed scheme.

\subsubsection{Formal security analysis}
The formal security analysis of CDAuth is as follows.

\textbf{Theorem 1:} Our CDAuth scheme is authentication-secure if no PPT adversary $\mathcal{A}$ has a nonnegligible advantage $Adv_{\mathcal{A}}^{CDAuth} (rt )$, in compromising CDAuth in the random oracle model (ROM).

To evaluate the authentication security of CDAuth, we adopt the DY adversarial model \cite{dolev2003security}. Let $\mathcal{A}$ be a polynomial-time adversary with runtime $rt$ in the ROM, where all entities access the collision-resistant, one-way hash functions $H_1(\cdot)$ and $H_2(\cdot)$. We define a random oracle query $H\vartheta$ such that $\mathcal{A}$ submits a message $m$ to the challenger $\mathcal{C}$, who returns a uniformly random value $\alpha$ and records the pair $(m, \alpha)$ in a maintained query log for subsequent reference.

\textbf{Proof.} The advantage of adversary $\mathcal{A}$ breaking the authentication security in CDAuth can be estimated as
\begin{equation}
\label{equ1}
\begin{aligned}
Adv_{\mathcal{A}}^{CDAuth} (rt ) &\le\frac{q_{hash}^{2}+(q_{exe}+q_{send})^2}{2|Hash|}\\&+ Adv_{\mathcal{A}}^{ECDLP}(rt)+2^{-|r|\lambda}+2\varepsilon.
\end{aligned}
\end{equation} 
Here, $q_{exe}$ represents the number of $Execute$ queries, in which the adversary $\mathcal{A}$ eavesdrops on communications to acquire public information. $q_{send}$ refers to the number of $Send$ queries, where the adversary  $\mathcal{A}$ manipulates messages to examine the protocol’s response to tampering. $q_{hash}$ denotes the number of hash queries performed as the adversary $\mathcal{A}$ attempts to find collisions or preimages to undermine security. $|Hash|$ specifies the size of the hash function’s output space, while $|r|$ indicates the cardinality of a subset of $[R]$, corresponding to a user’s attribute count. $Adv_{\mathcal{A}}^{ECDLP}(rt)$ quantifies the adversary’s advantage in solving the ECDLP within polynomial time $rt$, and $\varepsilon$ is a negligible value.

To prove (\ref{equ1}), we define four games, $Game_r$ for $r = 0, 1, 2, 3$. The adversary’s advantage in $Game_r$ is denoted by $Adv_{\mathcal{A}}^{Game_r}$, corresponding to the probability that $\mathcal{A}$ correctly predicts the challenge bit $c$ in that game.

$Game_0$: This game models an initial attack on CDAuth within the random oracle model, in which the adversary $\mathcal{A}$ tries to guess the authentication credentials. During the game, $\mathcal{A}$ issues $Test$ queries and receives the credentials whenever $c=1$. The semantic security of the credentials is defined as
\begin{equation}
\label{equ2}
    Adv_{\mathcal{A}}^{CDAuth} (rt )=|2Adv_{\mathcal{A}}^{Game_0}-1|.
\end{equation}

$Game_1$: This game simulates an eavesdropping attack by $\mathcal{A}$ using the $Execute$ query. $\mathcal{A}$ performs the $Test$ query to determine whether the output is the real credential $Aut_{m,i}$/$Aut_{Mm,i}$ between $ES_m$/$ES_M$ and $u_{m,i}$ or a random number. $\mathcal{A}$ also performs the $Test$ query to determine whether the output is the real credential $Aut_m$ between $ES_m$ and CS or a random number. In CDAuth, the credentials are calculated by $ES_m$/$ES_M$ and $u_{m,i}$ as $Aut_{m,i}  = \sum\nolimits_{j \in r}RR_{au_{A,j}} + RR_{au_{B}}$ or $Aut_{\!M\!m\!,i} \!\!=\!\! \sum\nolimits_{j \in r}\!R\!R\!B_{\!au_{\!A\!,j}}\!+\!R\!R\!B_{\!au_{\!B}}$, where $RR_{au_{A,j} }  = b_j^{ - 1} R_{au_{A,j} }' =\beta _j\! A_jg$, $RR_{au_{B} } = c^{ - 1} R_{au_{B} }'=\delta \zeta _{m,i} g$, $R\!R\!B_{au_{A\!,j} }\!\! = \!b_j^{ - 1}\!R_{au_{A\!,j} }''=\beta _j'\! A_jg$, $R\!R\!B_{au_{B} } \!\!=\! c^{ - 1}\!R_{au_{B} }''=\delta' \zeta _{m,i} g$. Similarly, the credential is also calculated by $ES_m$ and CS as $Aut_m  = H_2 (pid_m ||\eta _m )$, where $\eta _m  = \varphi _m /P\!K_{\!C\!S} sk_{es_m }$. Assumed that $\mathcal{A}$ can intercept all the messages $rf_{m,i}$, $uf_{m,i}$, $acf_{m,i}$ $V_{m,i}$ transmitted between $ES_m$ and $u_{m,i}$, $Req_M$, $Caf_{Mm,i}$, $Vb_{m,i}$ transmitted between $ES_M$ and $u_{m,i}$, $C\!om\!_M$, $C\!om\!_m$ transmitted between $ES_M$ and $ES_m$ and $Sig_m$, $cf_m$, $\hbar _m$ transmitted between CS and $ES_m$ over the common channel. But relying on these messages, it cannot derive the credentials $Aut_{m,i}$, $Aut_{Mm,i}$ and $Aut_m$. Therefore, the probability of $\mathcal{A}$ winning by eavesdropping attack in $Game_1$ will not increase. That is, adversary $\mathcal{A}$ has the same probability of winning $Game_0$ and $Game_1$, and hence
\begin{equation}
\label{equ3}
    Adv_{\mathcal{A}}^{Game_0}=Adv_{\mathcal{A}}^{Game_1}.
\end{equation}

$Game_2$: This game captures an active attack where $\mathcal{A}$ issues $Send$ and $H\vartheta$ queries and attempts to deceive participants into accepting forged messages. After intercepting $rf_{m,i}$, $uf_{m,i}$, $acf_{m,i}$ $V_{m,i}$, $Req_M$, $C\!om\!_M$, $C\!om\!_m$, $Caf_{Mm,i}$,  $Vb_{m,i}$, $Sig_m$, $cf_m$, and $\hbar _m$, $\mathcal{A}$ tries to modify these values to forge a valid message to pass verification. However, successful authentication requires knowledge of the secret keys $sk_{u_{m,i}}$, $sk_{es_m}$, and authentication credentials $Aut_{m,i}$, $Aut_{Mm,i}$, and $Aut_m$, all protected by an anti-collision one-way hash function. $\mathcal{A}$ repeats $H\vartheta$ query to find the conflict. Since each message is associated with the participant’s secret keys and random numbers, etc., according to the birthday paradox \cite{flajolet1992birthday}, we have
\begin{equation}
\label{equ4}
    |Adv_{\mathcal{A}}^{Game_2}-Adv_{\mathcal{A}}^{Game_1}| = \frac{q_{hash}^{2}+(q_{exe}+q_{send})^2}{2|Hash|}.
\end{equation}

$Game_3$: In this game, $\mathcal{A}$ tries to calculate the authentication credential $Aut_{m,i}$ or $Aut_{Mm,i}$ based on $rf_{m,i}$, $uf_{m,i}$, $acf_{m,i}$ $V_{m,i}$ or $Req_M$, $C\!om\!_M$, $C\!om\!_m$, $Caf_{Mm,i}$, $Vb_{m,i}$ intercepted in the common channel. To calculate $Aut_{m,i}  = \sum\nolimits_{j \in r}RR_{au_{A,j}} + RR_{au_{B}}$ or $Aut_{\!M\!m\!,i} \!\!=\!\! \sum\nolimits_{j \in r}\!R\!R\!B_{\!au_{\!A\!,j}}\!+\!R\!R\!B_{\!au_{\!B}}$, $\mathcal{A}$ need to calculate $RR_{au_{A,j} }$ from $RR_{au_{A,j} }  = b_j^{ - 1} R_{au_{A,j} }'$ and $RR_{au_{B} }$ from $RR_{au_{B} } = c^{ - 1} R_{au_{B} }'$ or $R\!R\!B_{\!au_{\!A\!,j}}$ from $R\!R\!B_{au_{A\!,j} }\!\! = \!b_j^{ - 1}\!R_{au_{A\!,j} }''$ and $R\!R\!B_{au_{B} }$ from $R\!R\!B_{au_{B} } \!\!=\! c^{ - 1}\!R_{au_{B} }''$. This is equivalent to requiring $\mathcal{A}$ to guess the random numbers $\{b_j\}_{j\in r}$ and $c$.  To calculate $Aut_m  = H_2 (pid_m ||\eta _m )$, $\mathcal{A}$ need to calculate $\eta _m$ from $\eta _m  = \varphi _m /P\!K_{\!C\!S} sk_{es_m }$, $\mathcal{A}$ need to calculate $sk_{es_m }$ from $pk_{es_m }$. This is equivalent to requiring $\mathcal{A}$ to solve ECDLP in polynomial time. Therefore,
\begin{equation}
\label{equ5}
    |Adv_{\mathcal{A}}^{Game_3}\!-Adv_{\mathcal{A}}^{Game_2}| \le Adv_{\mathcal{A}}^{ECDLP}(rt)+2^{-(|r|+1)\lambda}.
\end{equation}

Finally, the probability of $\mathcal{A}$ guessing $c = 1$ is
\begin{equation}
\label{equ6}
    |Adv_{\mathcal{A}}^{Game_3}|= \frac{1}{2}+\varepsilon.
\end{equation}

From (\ref{equ2}), (\ref{equ3}), and (\ref{equ6}), we have
\begin{equation}
\begin{split}
\label{equ7}
    \frac{1}{2}Adv_{\mathcal{A}}^{RMD-AC} (rt )&=|Adv_{\mathcal{A}}^{Game_0}- \frac{1}{2}|\\&=|Adv_{\mathcal{A}}^{Game_1}- \frac{1}{2}|\\&=|Adv_{\mathcal{A}}^{Game_2}- Adv_{\mathcal{A}}^{Game_3}+\varepsilon|.
\end{split}
\end{equation}

Based on (\ref{equ4}) and (\ref{equ5}), we can get
\begin{equation}
\begin{split}
\label{equ8}
Adv_{\mathcal{A}}^{Game_1}- Adv_{\mathcal{A}}^{Game_3} &= |Adv_{\mathcal{A}}^{Game_1}- Adv_{\mathcal{A}}^{Game_2}|\\&+|Adv_{\mathcal{A}}^{Game_2}- Adv_{\mathcal{A}}^{Game_3}|\\& \le \frac{q_{hash}^{2}+(q_{exe}+q_{send})^2}{2|Hash|}\\&+ Adv_{\mathcal{A}}^{ECDLP}(rt)+2^{-(|r|+1)\lambda}.
\end{split}
\end{equation}

Combining (\ref{equ7}) and (\ref{equ8}) leads to (\ref{equ1}), this completes the proof.

\subsubsection{Informal security analysis}
The informal security analysis of CDAuth is presented below.

(1) \textbf{Anonymity.} During edge server authentication, the CS assigns each ES a pseudo-ID $pid_m$ with an expiration time $ExpiryDate$ to hide its real identity. Similarly, intra-domain terminal authentication uses pseudo-IDs, with attributes and biometric data transmitted in hashed form $\{H_1(A_j)\}_{j \in r}$ and the biometric key $BIo{m,i}$, keeping real values secret. In cross-domain authentication, users rely on credential $Aut_{m,i}$ issued in their home domain $ES_m$. The cross-domain server $ES_M$ can only access the group-operated attributes $A_jg = C_jks_{m-M}^{-1}$ after decryption. Recovering the original attributes requires breaking the session key and solving the ECDLP, which is computationally infeasible. Thus, the scheme preserves anonymity of identities, attributes, and biometrics. 

(2) \textbf{Defending impersonation attack.} In the edge server authentication phase, an adversary attempting impersonation must forge a valid signature $Sig_m = (pid_m, ts_m, F_m, sig_m)$, where $F_m = f_m g$, $sig_m = f_m + e_{pid_m}\varpi_m$, and $\varpi_m = H_2(pid_m | ts_m | T_{pid_m})$. Since the secret $f_m$ is never revealed, constructing $(F_m, sig_m)$ is infeasible. In the terminal intra-domain phase, impersonation requires regenerating the registration factor $\!r\!f_{\!m,i}\!\!= (id_{u_{m,i}},\!ts_{u_{m,i}},\{H_1\!(A_j)\}_{j\in r},B\!I\!o_{\!m,i},\!U\!V\!e_{\!m,i})$, where $BIo_{m,i} = H_1(\zeta_{m,i})$, $Vt_{m,i} = H_1(H_1(A_j)||BIo_{m,i})$, and $UVe_{m,i} = id_{u_{m,i}} ts_{u_{m,i}} Vt_{m,i}sk_{u_{m,i}} pk_{es_m}$. These computations involve $A_j$, $Bio_l$, and $sk_{u_{m,i}}$, which remain secret, preventing forgery. Furthermore, cross-domain authentication requires a valid intra-domain credential. As the adversary cannot succeed in impersonation during the intra-domain phase, it is also unable to construct a valid cross-domain request. Therefore, the proposed scheme effectively prevents impersonation across all authentication phases.

(3) \textbf{Defending Man-in-the-Middle attack.} Suppose that an adversary $\mathcal{A}$ attempts a man-in-the-middle attack by impersonating either ESs or users to transmit forged messages to the legitimate communication partner. In CDAuth, all received messages and sender states are verified, which is infeasible without access to secret information. For example, in the cross-domain authentication phase, a user $u_{m,i}$ sends an access request $Req_M = (pid_m, id_{u_{m,i}}, VA_{m,i}, AAu_{u_{m,i}}, R_{au_{A,j}}, R_{au_B})$ to $ES_M$ for cross-domain authentication. To impersonate the user, $\mathcal{A}$ would need to generate a valid $Req_M$ that can pass verification. This requires knowledge of the authentication credential $Aut_{m,i}$, which is protected by the collision-resistant one-way hash function, as well as the private key $sk_{u_{m,i}}$. Since $Aut_{m,i}$ is generated via group operations and private randomness, and recovering $sk_{u_{m,i}}$ is computationally equivalent to solving the ECDLP, constructing a valid forgery is infeasible. Hence, CDAuth effectively resists man-in-the-middle attacks.

(4) \textbf{Defending replay attack.} In CDAuth, the terminals and ESs employ pseudo-identities $id_{u_{m,i}}$ and $pid_m$ with expiration times $ExpiryDate$, ensuring their validity only within the designated period. Additionally, the incorporation of nonces and timestamps into message exchanges guarantees freshness, thereby providing effective resistance against replay attacks.

(5) \textbf{Defending collusion attack.} In the intra-domain authentication phase, two users may attempt to collude to obtain a credential for an attribute set exceeding their individual permissions. For example, user $u_{m,i}$ holds ${A_1, A_2}$, and user $u_{m,i+1}$ holds ${A_1, A_3}$; together they may try to obtain a credential for ${A_1, A_2, A_3}$. Specifically, $u_{m,i}$ executes the $\textbf{RegGen}$ algorithm to generate the registration factor $\!r\!f_{\!m,i}\!\!= (id_{u_{m,i}},ts_{u_{m,i}},\{H_1\!(A_1),H_1\!(A_2),H_1\!(A_3)\},B\!I\!o_{\!m,i},\!U\!V\!e_{\!m,i})$ and submits it to $ES_m$. The $\textbf{VerReg}$ algorithm at $ES_m$ detects that the combined attribute set ${A_1, A_2, A_3}$ does not match $u_{m,i}$’s legitimate attributes, aborting authentication. Similarly, in cross-domain authentication, $ES_m$ executes Step 4 of the $\textbf{MutESCom}$ algorithm to validate the identity of $u_{m,i}$ using $H\!A_M\!R\!Au_{m\!,i}' \!\!= \!\!pk_{u_{m\!,i} }\!V\!A_{m\!,i}$, where $R\!Au_{m\!,i}'\!=\!pid_m\!H_2 (R_{\!au_{A\!,1} }\!||R_{\!au_{A\!,2} }||\!R_{au_{\!B} } )$, $VA_{m\!,i} \!\!=\!\! H_2(Aut_{m\!,i})\!H_2 (R_{au_{A\!,1} }||R_{au_{A\!,2} }||R_{au_{A\!,3} }||R_{au_{B} })pid_m sk_{u_{m,i}}^{-1}$. Since the equality does not hold, the verification fails, and the cross-domain authentication is terminated. Therefore, the proposed scheme is resilient against user collusion attacks.

\section{Performance analysis}
This section analyzes the performance of CDAuth. All algorithms are implemented in Java using the JPBC library and tested on a MacBook Pro (13-inch, M2, 2022) running macOS 12.5. SHA-256 and AES (Advanced Encryption Standard) are adopted as the hash and symmetric encryption primitives, respectively, and the corresponding results are reported in Table \ref{tab1}. Moreover, the Fuzzy Extractor is implemented in Python to measure the costs of generating and recovering 256-bit biometric keys, with the results also included in Table \ref{tab1}.

\begin{table}[htbp]
\caption{Execution time of each algorithm}
\vspace{-10pt}
\begin{center}
\scalebox{0.95}{
\begin{tabular}{ c c }
\hline
\textbf{The algorithm} & \textbf{Execution time (ms)} \\
\hline
Multiplication operation $(T_m )$ & $T_m  \approx 0.0003$ \\
Addition operation $(T_a)$ & $T_a \approx 0.0001$ \\
Exponentiation operation in $G$ $(T_e )$ & $T_e  \approx 6.6038$ \\
Exponentiation operation in $G_T$ $(T_{pe } )$ & $T_{pe }  \approx 4.9873$ \\
Elliptic curve point addition $(T_{pa } )$ & $T_{pa}  \approx 0.0281$ \\
Elliptic curve point multiplication $(T_{pm} )$ & $T_{pm}  \approx 6.479$ \\
Hash operation $(T_h )$ & $T_h  \approx 0.0001$ \\
Bilinear pairing $(T_{bp} )$ & $T_{bp}  \approx 4.4597$ \\
Bilinear pairing scalar multiplication $(T_{sm} )$ & $T_{sm}  \approx 5.0819$ \\
XOR operation $(T_X)$ & $T_X \approx 0.0108$\\
Symmetric encryption ($T_{Enc}$) & $T_{Enc}  \approx 2.48$ \\
Symmetric decryption ($T_{Dec}$) & $T_{Dec}  \approx 0.461$ \\
Biological key generation $(T_{Gen})$ & $T_{Gen} \approx 4.9222$ \\
Biological key recovery $(T_{Rep})$ & $T_{Rep} \approx 0.0172$ \\
\hline
\end{tabular}
}
\label{tab1}
\end{center}

\vspace{-12pt}
\end{table}

To our knowledge, no existing cross-domain authentication scheme simultaneously integrates ID, attributes, and biometrics. We therefore evaluate the computational complexity of our scheme in comparison with He \emph{et al.}'s \cite{11048743}, Huang \emph{et al.}'s \cite{huang2022blockchain}, Zhang \emph{et al.}'s \cite{9785665}, Tian \emph{et al.}'s \cite{tian2024identity} and Shukla \emph{et al.}'s \cite{shukla2022novel} schemes, as summarized in Table \ref{tab3}. He \emph{et al.} \cite{11048743} and Huang \emph{et al.} \cite{huang2022blockchain} present attribute-based approaches but lack explicit cross-domain procedures; thus, their complexities are estimated from intra-domain authentication while ignoring blockchain queries and read/write operations. Zhang \emph{et al.} \cite{9785665} employ ID, PUF keys, and biometrics for multi-factor cross-domain authentication, whereas PUF and accumulator costs are not considered. Tian \emph{et al.}'s \cite{tian2024identity} and Shukla \emph{et al.}'s \cite{shukla2022novel} schemes focus on multi-factor authentication using ID, passwords, and biometrics without supporting cross-domain scenarios. Zero-knowledge proof operations of these schemes and our computations related to transmission security during transmission are also omitted in complexity calculations.

\begin{table*}[htbp]
\caption{Computational complexity of our scheme and the other schemes}
\begin{center}
\begin{tabular}{|c|c|c|}
\hline
 & Intra-domain authentication & Cross-domain Authentication  \\
\hline
He \emph{et al}. \cite{11048743} (attribute-based) & $\makecell[c]{3|r|T_{bp} + (3|r| + 4)T_e + (|r|+2)T_{pm} 
  + 2T_h  \\+ |r|T_{pa}+T_{m} \approx 39.6976|r|+39.3737 \: \rm{ms}}$ & $\begin{array}{c}
 (|r| + 2)T_e + T_{pm} 
  + T_h  + |r|T_{pa}+T_{m}\\\approx 6.6319|r|+19.687 \: \rm{ms}
 \end{array}$ \\
\hline
Huang \emph{et al}. \cite{huang2022blockchain} (attribute-based) & $\begin{array}{c}
 \!(7|r|\! + \!12)T_e \!+(2|r|+5)T_{pe}\!+\!(9\! + \!2|r|)T_{bp}  
 \! + \!(2|r|+5)T_{sm} \! \\+ \!(7|r|+1)T_{pa}\!+\!|r|T_m \!+ \!T_h \approx 75.4814|r|\!+\!169.7574 \: \rm{ms}
 \end{array}$ & $\begin{array}{c}
 \!(6|r|\! + \!7)T_e \!+ (|r|+3)T_{pe}\!+\!(5\! + \!|r|)T_{bp}  
 \! + \\\!(|r|+3)T_{sm} \! \!+ \!6|r|T_{pa}\!+\!|r|T_m \!+ \!T_h \\\approx 54.3206|r|+98.7328 \: \rm{ms}
 \end{array}$ \\
\hline
Zhang \emph{et al}. \cite{9785665} (multi-factor) & $\begin{array}{c}
 4T_{pm} \! +\!5 T_{a} + \!5 T_{m} \!+ \!9T_h+T_{Enc}+T_{Dec}+T_{Gen} \\ \approx 33.7821 \: \rm{ms}
 \end{array}$ &  $\begin{array}{c}
 5T_{pm} \!+\!5T_{a} + \!6 T_{m} \!+ \!13T_h+2T_{Enc}+\\2T_e+T_{Dec}+T_{Gen}  \approx 55.9494 \: \rm{ms}
 \end{array}$ \\
\hline
Tian \emph{et al}.  \cite{tian2024identity}(multi-factor) & $\begin{array}{c}
 \!  3T_{pm} \! + \!T_{a} + \!T_{m} \!+ \!6T_h+T_{bp}+T_{pe}+T_{pa}+3T_{X}+\\T_{Gen}+T_{Rep}  \approx 33.8849 \: \rm{ms}
 \end{array}$ &  $-$ \\
\hline
Shukla \emph{et al}. \cite{shukla2022novel} (multi-factor) & $\begin{array}{c}
 \!  5T_{pm} \!+ \!19T_h+T_{pa}+3T_{X}+T_{Gen}+T_{Rep}+T_{Enc} \\\approx 39.8768 \: \rm{ms}
 \end{array}$ &  $-$ \\
\hline
Ours (CDAuth) & $\begin{array}{c}
 (2|r|+2)T_{pm}  
  +(|r|+3)T_h  + |r|T_{a} +|r|T_{pa}+T_{Gen}\\+(|r|+2)T_{m} \approx 12.9866|r|+17.8811 \: \rm{ms}
 \end{array}$ & $\begin{array}{c}
 (|r| + 1)T_{pm} 
  + 2T_h  + |r|T_{pa}+2T_{m}\\\approx 6.5071|r|+6.4798 \: \rm{ms} \\ 
 \end{array}$\\
\hline

\end{tabular}
\label{tab3}
\end{center}
\vspace{-5pt}
\end{table*}

It can be seen from Table \ref{tab3} that, among attribute-based schemes, our scheme achieves the lowest computational complexity in intra-domain authentication, followed by He \emph{et al.}'s scheme \cite{11048743}, whereas Huang \emph{et al.}'s scheme \cite{huang2022blockchain} incurs the highest cost. In the cross-domain phase, our scheme exhibits a complexity close to but slightly lower than that of He \emph{et al.}'s scheme \cite{11048743}, and remains more efficient than Huang \emph{et al.}'s scheme \cite{huang2022blockchain}. When compared to multi-factor authentication schemes, Shukla \emph{et al.}'s scheme \cite{shukla2022novel} has the highest complexity, while Zhang \emph{et al.}'s \cite{9785665} and Tian \emph{et al.}'s \cite{tian2024identity} schemes show similar complexity, with Zhang \emph{et al.}'s scheme \cite{9785665} being slightly lower. Our scheme consistently demonstrates the lowest computational complexity.

\begin{figure*}
 \centering
 \begin{minipage}[t]{0.24\linewidth}
  \centering
  \includegraphics[width=\linewidth]{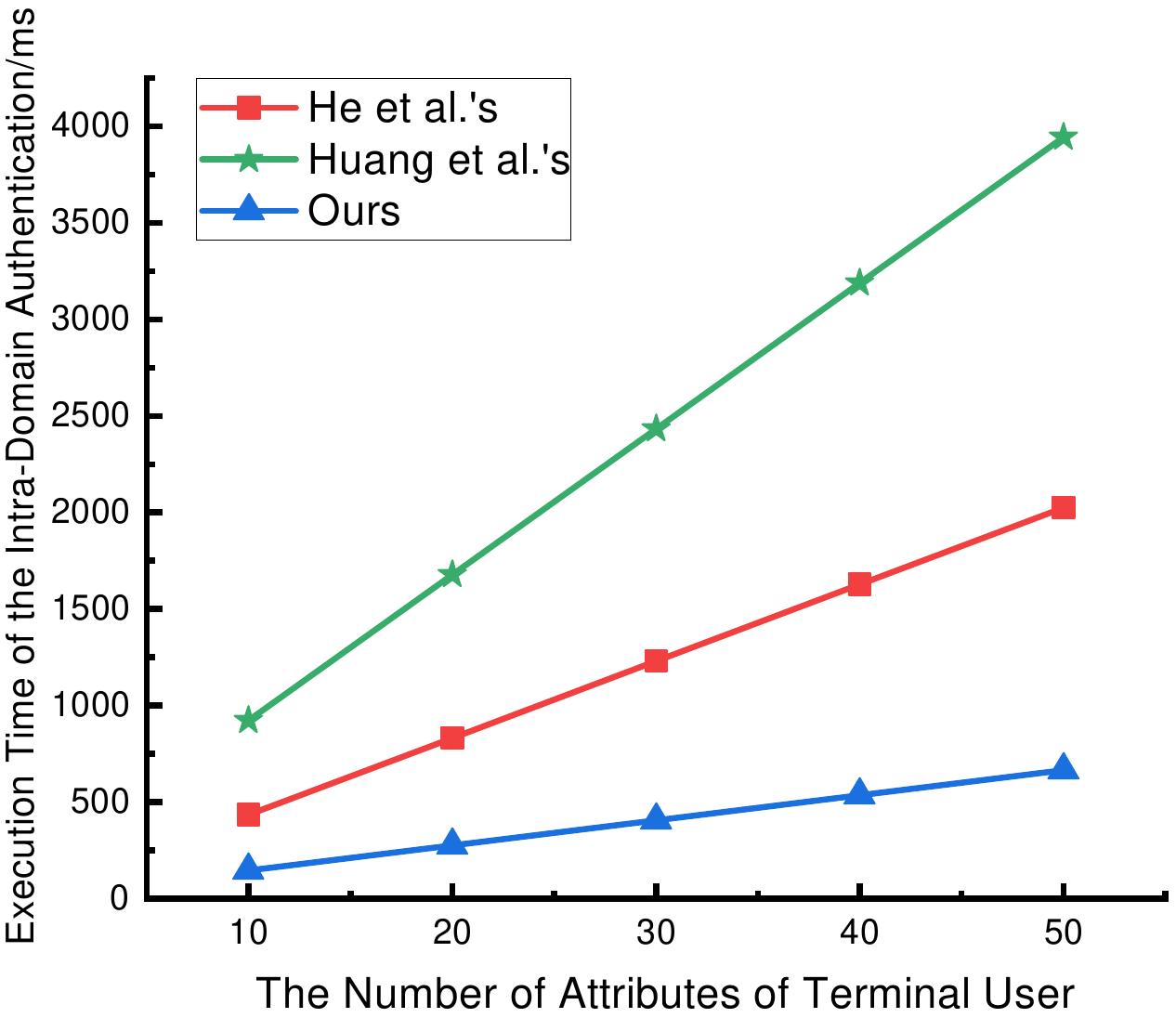}
    \textbf{(a)} \footnotesize\textsf{Intra-domain (Attribute)}
 \end{minipage}~
 \begin{minipage}[t]{0.24\linewidth}
  \centering
  \includegraphics[width=\linewidth]{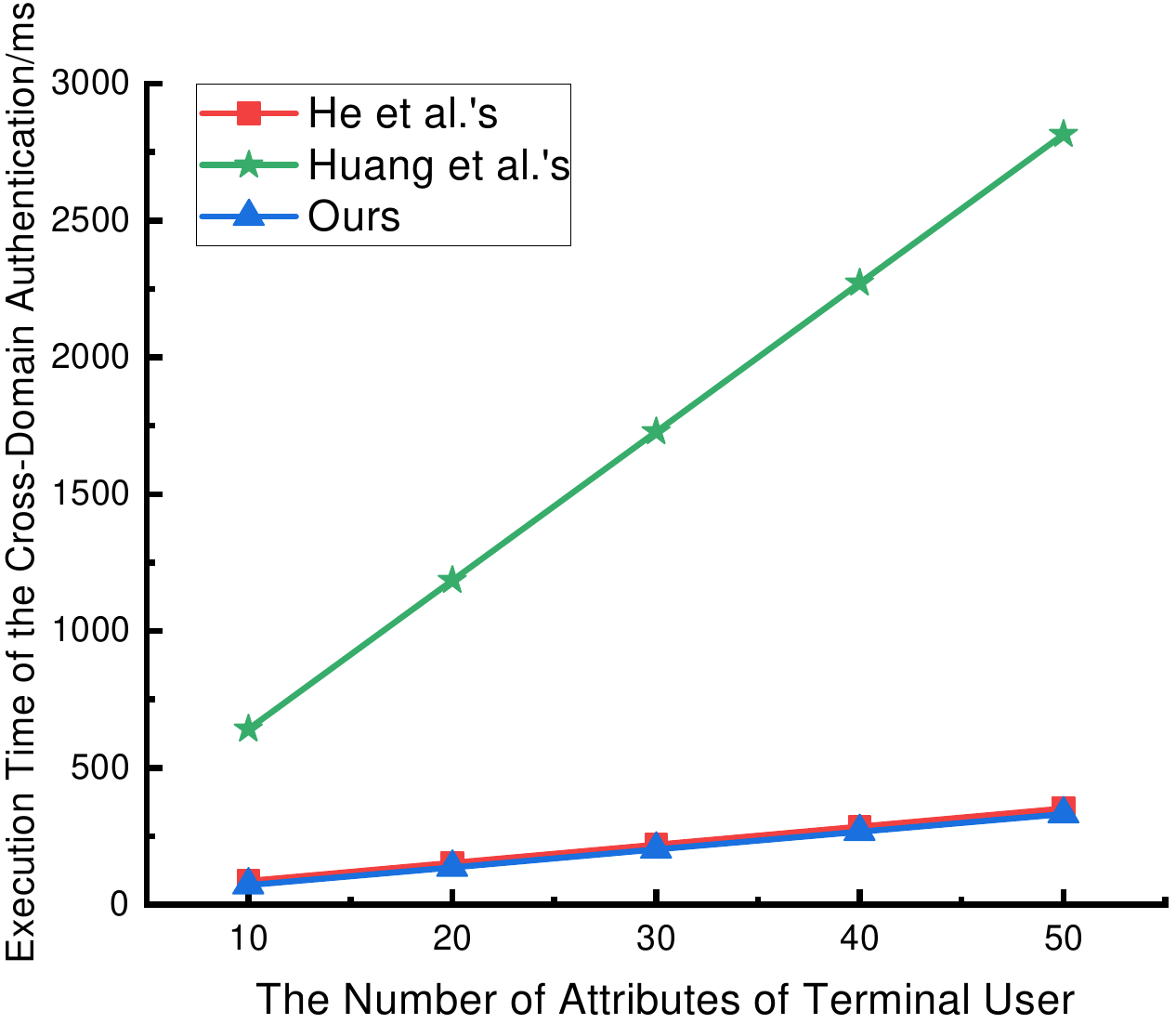}
  \textbf{(b)} \footnotesize\textsf{Cross-domain (Attribute)}
 \end{minipage}~
\begin{minipage}[t]{0.24\linewidth}
  \centering
  \includegraphics[width=\linewidth]{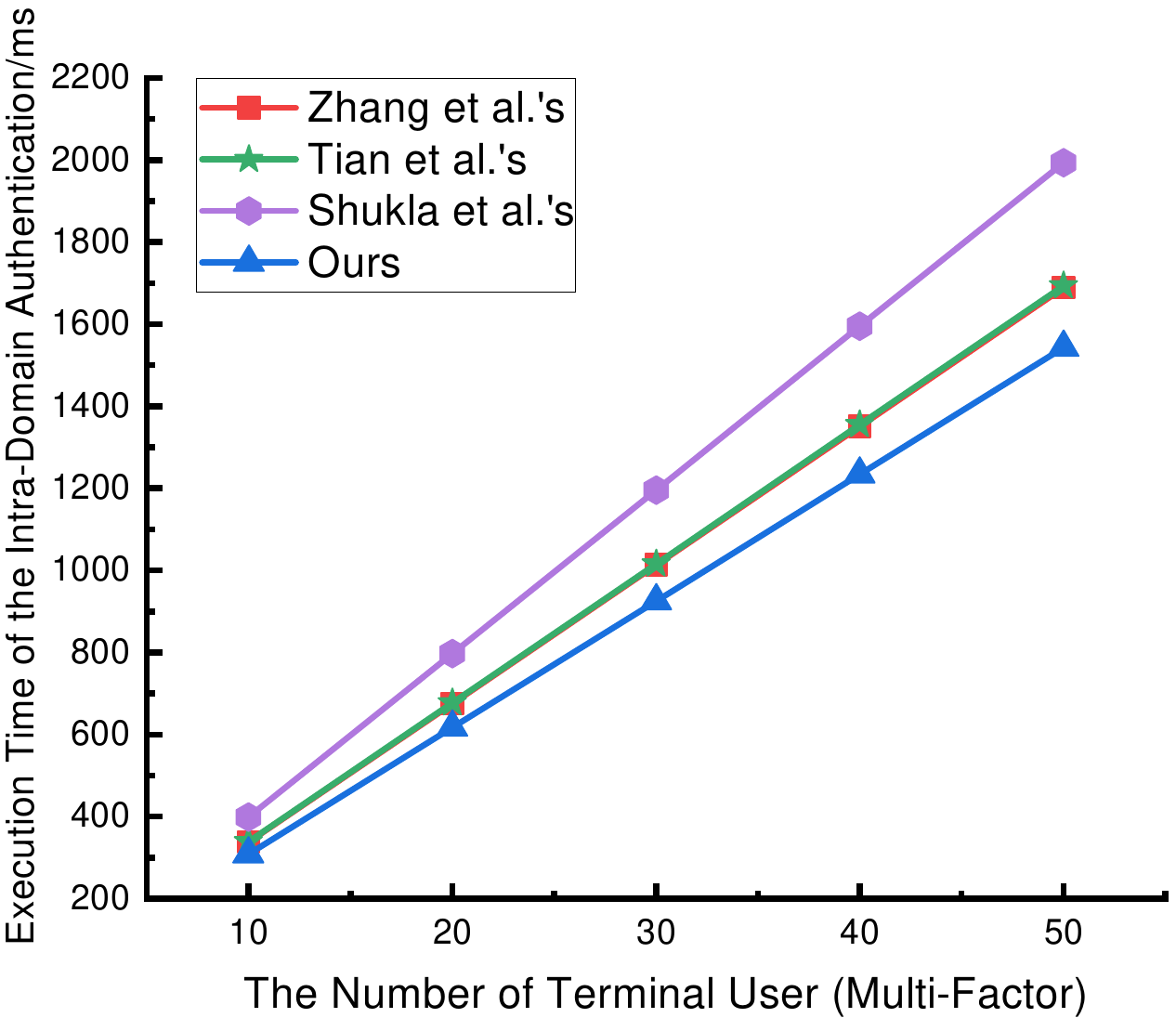}
  \textbf{(c)} \footnotesize\textsf{Intra-domain (Multi-factor)}
 \end{minipage}~
 \begin{minipage}[t]{0.24\linewidth}
  \centering
  \includegraphics[width=\linewidth]{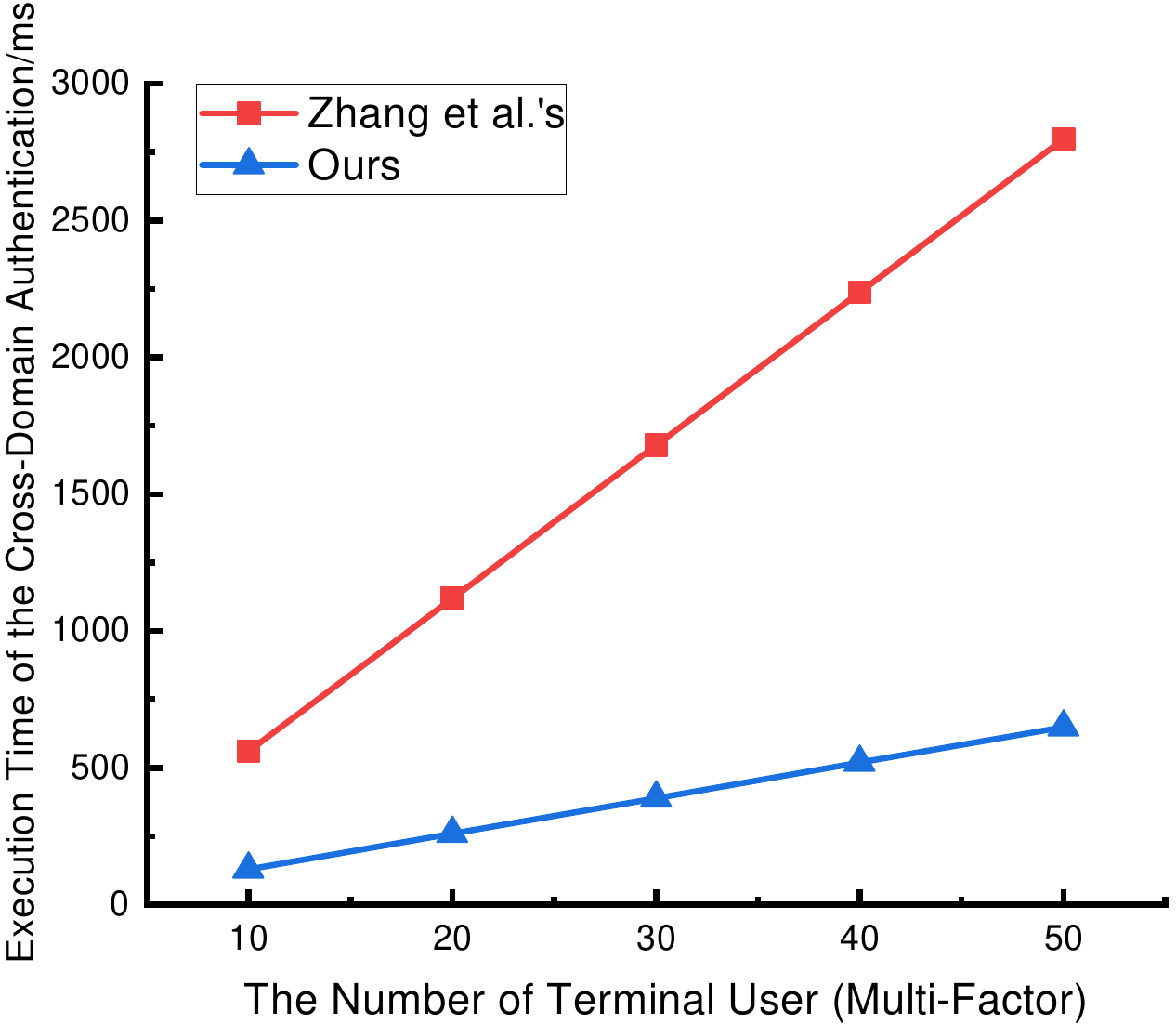}
  \textbf{(d)} \footnotesize$\textsf{Cross-domain (Multi-factor)}$
 \end{minipage}~

 \caption{Comparison of computation cost}
 \label{example}
 \vspace{-15pt}
\end{figure*}

Then, we conduct data simulations to evaluate the performance of CDAuth. Using the data in Table \ref{tab1}, we calculate and visualize the computational costs of CDAuth and the schemes of He \emph{et al.} \cite{11048743}, Huang \emph{et al.} \cite{huang2022blockchain}, Zhang \emph{et al.} \cite{9785665}, Tian \emph{et al.} \cite{tian2024identity}, and Shukla \emph{et al.} \cite{shukla2022novel}. For attribute sizes $|r| \in \{10,20,30,40,50\}$, we analyze intra-domain and cross-domain costs on the user side against He \emph{et al.}'s \cite{11048743} and Huang \emph{et al.}'s \cite{huang2022blockchain} schemes. For $n \in \{10,20,30,40,50\}$ users with three fixed authentication factors, we compare intra-domain costs with Zhang \emph{et al.}'s \cite{9785665}, Tian \emph{et al.}'s \cite{tian2024identity}, and Shukla \emph{et al.}'s \cite{shukla2022novel} schemes, including cross-domain costs for Zhang \emph{et al.}'s \cite{9785665}. The results are shown in Figure \ref{example}.

As shown in Figure \ref{example}-\textbf{(a)}, the intra-domain authentication time of each user increases linearly with $|r|$. Under the same number of attributes—where our scheme employs ID, attributes, and biometrics while concealing actual information via group operations—Huang \emph{et al.}’s scheme \cite{huang2022blockchain} incurs the highest cost, followed by He \emph{et al.}’s scheme \cite{11048743}, whereas our scheme demonstrates significantly lower costs.

Figure \ref{example}-\textbf{(b)} illustrates the computational costs of attribute-based cross-domain authentication across the three protocols. For the same number of attributes, Huang \emph{et al.}’s scheme \cite{huang2022blockchain} incurs the highest computational overhead, while our scheme and He \emph{et al.}’s scheme \cite{11048743} exhibit comparable performance, with our scheme requiring slightly lower costs.

When comparing with the three multi-factor authentication schemes, we set the number of authentication factors to three ($|r|=1$ in CDAuth) for consistency. Figure \ref{example}-\textbf{(c)} shows that computational cost grows linearly with the number of users. Shukla \emph{et al.}’s scheme \cite{shukla2022novel} exhibits the highest cost, Tian \emph{et al.}'s  \cite{tian2024identity} and Zhang \emph{et al.}'s \cite{9785665} schemes show similar performance (Zhang \emph{et al.} slightly lower), while our scheme consistently requires less computation. Unlike existing multi-factor authentication schemes, which are hardware-assisted and ECC-based to achieve lightweight computation, our scheme does not rely on hardware acceleration but supports fine-grained access control, introducing modest overhead while achieving better performance and broader functionality.

To our knowledge, only a few multi-factor schemes support cross-domain authentication, with Zhang \emph{et al.}’s scheme \cite{9785665} selected for comparison. As shown in Figure \ref{example}-\textbf{(d)}, in the cross-domain phase, for the same three authentication factors and the same number of users, our scheme requires lower computational costs than that of Zhang \emph{et al.}'s scheme \cite{9785665}.

\section{Conclusion}

Aiming at the frequent cross-domain interactions of terminal users in complex network environments, we propose a cross-domain multi-factor authentication protocol based on edge-cloud collaboration. By offloading part of the authentication workload to nearby edge servers, the protocol reduces communication latency while the decentralized architecture alleviates single-point-of-failure and key escrow issues of centralized authentication. During authentication, attributes and biometrics are encrypted and hidden, with the ciphertext used for authentication. This approach achieves efficient, flexible, and secure authentication,  laying the foundation for fine-grained access control. Security and performance evaluations demonstrate that the proposed protocol achieves superior security and efficiency. In future work, we will extend the protocol to strengthen its resist stronger adversarial models, such as the CK, eCK model and key-compromise impersonation (KCI) attacks, and explore distributed authentication mechanisms.

\section*{Acknowledgment}
This work is supported by the National Science and Technology Major Project (2025ZD1607700), Natural Science Foundation of Shanghai under Grant 25ZR1402098, Shanghai Science and Technology project under Grant 24LZ1401600, the startup fund from East China University of Science and Technology under Grant YH0142234.

\bibliographystyle{IEEEtran}
\bibliography{IEEEabrv,deraining}

\begin{IEEEbiography}[{\includephoto{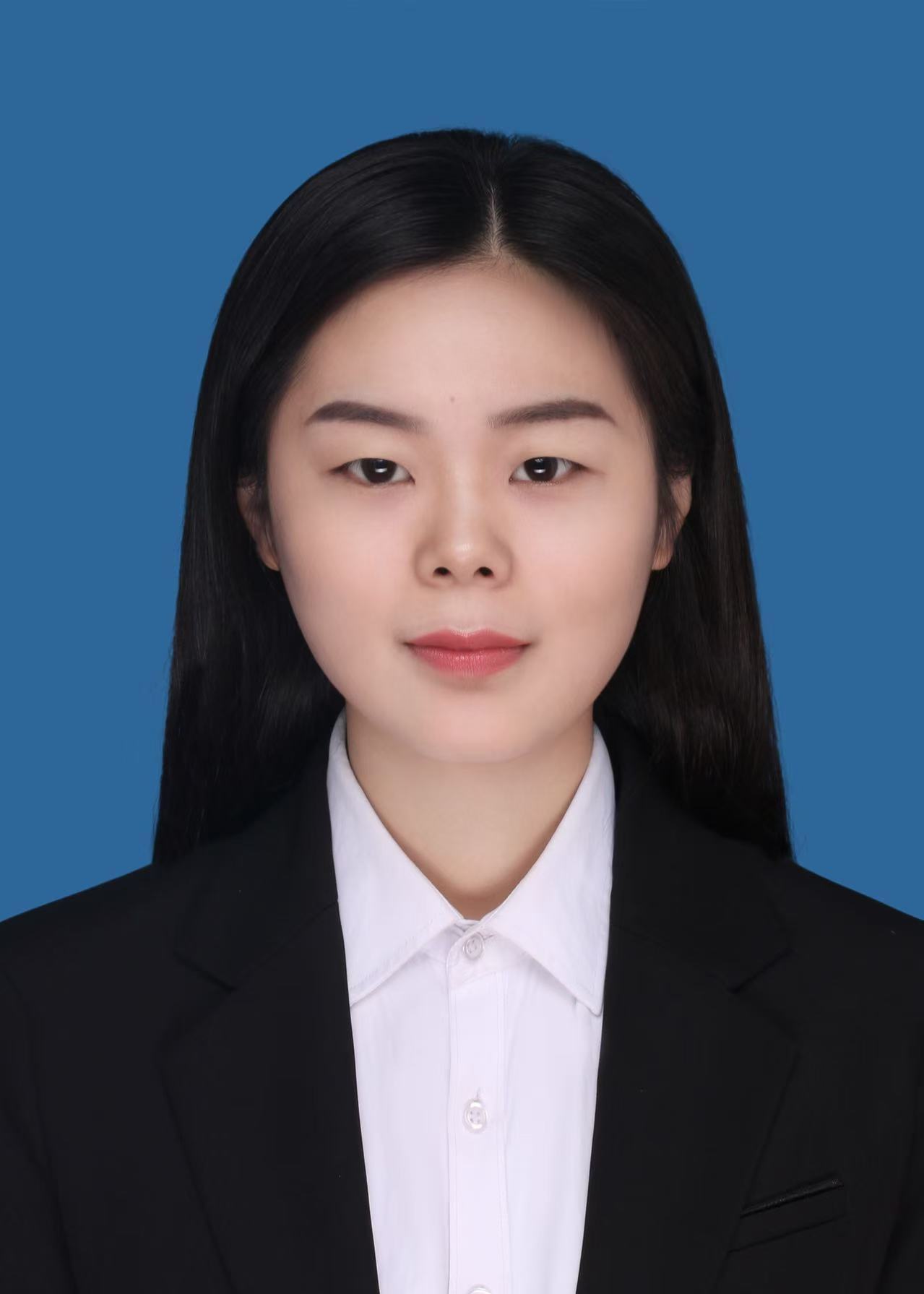}}]{Yongjiao Li} is currently pursuing her Ph.D. degree at the School of Information Science and Engineering, East China University of Science and Technology, Shanghai, China. Her current research interests include Internet of Things security, attribute-based encryption and applied cryptography.
\end{IEEEbiography}

\begin{IEEEbiography}[{\includephoto{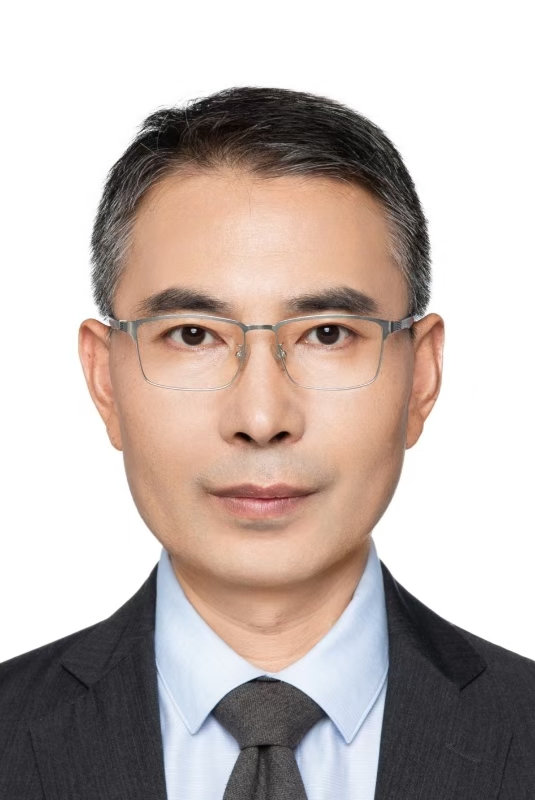}}]{Zhenlei Wang} received the B.S. and Ph.D. degrees in control science and engineering from Northeastern University, Shenyang, China, in 1997 and 2002, respectively. He is currently a Professor with the School of Information Science and Engineering, East China University of Science and Technology and also with the Key Laboratory of Smart Manufacturing in Energy Chemical Process, Ministry of Education. His research interests include intelligent control, modeling and analysis the characteristic of complex systems, intelligent optimization algorithms and fault diagnosis.
\end{IEEEbiography}

\begin{IEEEbiography}[{\includephoto{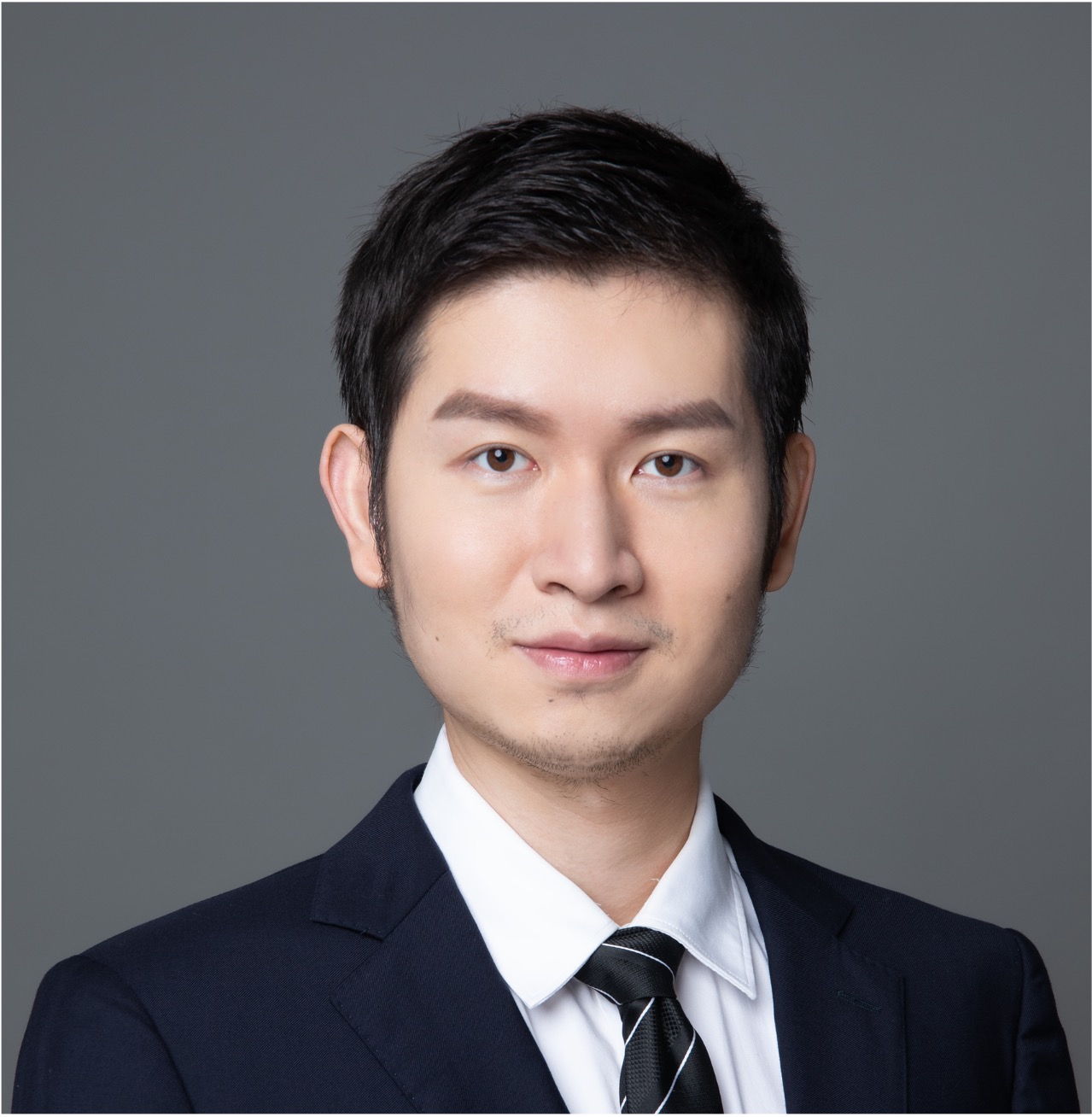}}]{Zhu Cao} received the B.Sc. and Ph.D. degrees in computer science and technology from Tsinghua University, Beijing, China, in 2013 and 2017, respectively. He is currently an Associate Professor with College of Electronics and Information Engineering, Tongji University, and with Shanghai Research Institute for Intelligent Autonomous Systems, Tongji University. His research interests include information security, quantum information, and artificial intelligence.
\end{IEEEbiography}

\vfill

\end{document}